\documentclass[12pt,a4paper]{article}
\usepackage{graphicx}
\usepackage{latexsym}
\title{On modelling transitional turbulent flows
using under-resolved direct numerical simulations:\\
The case of plane Couette flow}
\author{Paul Manneville and Joran Rolland\\
Laboratoire d'Hydrodynamique\\
\'Ecole Polytechnique, F-91128 Palaiseau, France}

\date{Accepted on August 24, 2010, to appear in TCFD}
\begin{document}

\sloppy

\maketitle

\begin{abstract}
Direct numerical simulations have proven of inestimable help to our understanding of the transition to turbulence in wall-bounded flows.
While the dynamics of the transition from laminar flow to turbulence
via localised spots can be investigated with reasonable computing resources
in domains of limited extent, the study of the decay of turbulence
in conditions approaching those in the laboratory requires consideration
of domains so wide as to
exclude the recourse to fully resolved simulations. Using Gibson's
\textsf{C++} code \textsc{ChannelFlow}, we scrutinize the effects of a
controlled lowering of the numerical resolution on the decay of
turbulence in plane Couette flow at a quantitative level. We show
that the number of Chebyshev polynomials describing the cross-stream
dependence can be drastically decreased while preserving all the
qualitative features of the solution. In particular, the oblique
turbulent band regime experimentally observed in the upper part of
the transitional range is extremely robust. In terms of
Reynolds numbers, the resolution lowering is seen to yield a regular
downward shift of the upper and lower thresholds $R_{\rm t}$ and
$R_{\rm g}$ where the bands appear and break down. The study is
illustrated with the results of two preliminary experiments.
\end{abstract}

\section{Introduction}
\label{s1}

The `laminar--turbulent' transition in globally subcritical flows is far
from being fully understood. This is due to its abrupt and hysteretic
nature, and to the fact that phase space coexistence, typical of
a subcritical bifurcation, has a nontrivial counterpart in physical space,
with laminar flow and turbulence coexisting in separate regions of the
flow domain. Here we focus on plane Couette flow (PCF), the flow of a
viscous fluid with kinematic viscosity $\nu$ sheared between two parallel
plates at a distance $2h$, translating in opposite directions at speeds
$\pm U$. This flow configuration is free from global advection. The
{\it laminar flow\/} is known to be linearly stable for all values of
the Reynolds number $R=Uh/\nu$, whereas under usual conditions
{\it turbulent\/} flow takes place for $R$ large enough, typically
$R\sim400$.

In fact the transition can be examined in both directions,
`laminar$\to$turbulent' (direct) and `turbulent$\to$laminar' (reverse).
Many early studies have dealt with the direct transition, and especially
with the dynamics of {\it turbulent spots\/}, by means of laboratory
experiments or numerical simulations.
More recently, experiments performed at Saclay~\cite{Petal03} have shown
that the reverse transition is marked by the occurrence of {\it oblique
turbulent bands\/}, only observable in very large aspect ratio%
\footnote{The aspect ratio is the dimensionless size of the set-up, i.e.
its lateral size in units of the cross-stream half-gap $h$.\label{fn1}}
systems, in some range
$R_{\rm g}<R<R_{\rm t}$.
In the lowest part of this range,%
\footnote{Subscript `g' used hereafter stands for `global' in the sense of
`global stability threshold', the value of $R$ below which the laminar
base flow profile is unconditionally recovered in the long term, i.e.,
whatever the strength of the perturbation brought to the flow.}
near $R_{\rm g}\simeq325$, the turbulent bands become fragmented and turn
into spots of irregular shape before decaying after long transients when
$R$ is further decreased below $R_{\rm g}$. Hysteresis is observed and
sustained turbulent spots
can be obtained by triggering the laminar flow with sufficiently large
local perturbations when $R>R_{\rm g}$, whereas the laminar profile can
be maintained up to much higher values of $R$ provided that the experiment
is sufficiently clean. At the upper end of the transitional range,
the pattern disappears progressively and the transition from the
turbulent bands to {\it featureless turbulence\/} at
$R_{\rm t}\simeq410$ is continuous. The term `featureless' used here
is borrowed from \cite{Aetal86} where it served  to describe the
high-$R$ turbulent regime beyond spiral turbulence in
Taylor--Couette flow which corresponds to the oblique turbulent band
pattern in PCF.

Besides laboratory experiments, numerical simulations of the
Navier--Stokes equations (NSE) have provided invaluable information. 
An important output of early computations was the concept of
minimal flow unit (MFU) of size just necessary to maintain
turbulence in a wall-bounded flow \cite{JM91}, a fundamental ingredient
in the elucidation of the mechanisms sustaining turbulence \cite{HKW95}.
Later the MFU context was extensively used to study the decay of
turbulence within the framework of dynamical systems theory
\cite{Eetal08}. Simultaneously, numerical simulations were also
performed in wider domains, which lead to the discovery of a large
scale streamwise structures in turbulent PCF \cite{Ketal96} and
other wall-bounded flows at Reynolds number somewhat beyond the
transitional range defined above \cite{ptrsa07}. 

Numerical studies specially dedicated to the problem of oblique turbulent
bands are recent. Soon after the experiments that put them at the
forefront, Barkley \& Tuckerman \cite{BT05-07} succeeded in reproducing
the fact by simulating the NSE in domains elongated in the expected direction of the pattern's wavevector but narrow in the complementary
in-plane direction. These simulations gave useful information on the
pattern, properly accounting for the essential features of the
laminar-turbulent alternation. The mechanism producing the bands has
however remained elusive up to now, and it is not clear whether
periodic boundary conditions a few MFUs apart along the short dimension
of these domains do not handicap our understanding of it.
Although the occurrence of bands appears to be an extremely robust phenomenon, as our study will confirm, it thus seemed interesting to
consider cases where the long-range streamwise coherence of the
large scale streaky structures commonly observed in wall-bounded
flows~\cite{ptrsa07} was sufficiently well embraced. The coherence
length of these structures being indeed at least one order of magnitude
larger than the streamwise length of the MFU, this revives simulations
in large aspect ratio, conventionally-oriented,
domains. Such simulations again showed the occurrence of oblique
turbulent bands~\cite{Detal10,Tetal09}.

The computationally demanding character of these fully resolved numerical
experiments calls for the exploration of alternate approaches involving
some more or less well controlled level of approximation. This
perspective was taken in \cite{LM07} where the flow was modelled
using a Galerkin expansion of the NSE in the
cross-stream direction $y$ in terms of well-chosen {\it ad hoc\/}
polynomials. The main characteristics of the transition were recovered at
much lower numerical cost from a truncation of the expansion at first
significant order, which permitted simulations in very large aspect-ratio
domains \cite{Ma09}. However, the transitional range was lowered by a
factor of two with respect to the experiments as a result of insufficient
energy transfer and dissipation in the cross-stream direction.
Furthermore the oblique bands in the upper part of the transitional range
were not obtained, presumably another effect of the lowered cross-stream
resolution.

The purpose of the work presented here is not
to improve the model mentioned above by truncating it at higher orders,
which is possible but very cumbersome and opaque, but to test this
resolution effect in the context of direct numerical simulations,
thus considering the deliberate decrease of the spatiotemporal
resolution as a systematic modelling strategy. Our motivation is basically
that, since qualitative and quantitative comparisons of solutions obtained
at different resolutions are easy, the degree of approximation can be
evaluated with some confidence. Having tested the reliability of this
procedure, we may expect to obtain clues on the mechanisms of band
formation and decay directly from the NSE at reduced numerical cost,
in much the same way as lowering the size of the computational domain
down to the dimensions of the MFU has helped toward the understanding of
the self-sustaining process~\cite{Wa97}. Finally, if quantitatively
reliable low resolution simulations can be performed, studying the
statistics of the upper transition at $R_{\rm t}$ as well as the lower
transition at $R_{\rm g}$ will be possible in larger domains, during
longer periods of time (the so-called `thermodynamic limit' involved in
analogies with thermodynamic phase transitions \cite{Po86}), which will
go in the same direction as in \cite{Ma09} but without the limitations of
the model used in that work.
Encouragement to follow the program sketched above can also be found
in the work of Willis \& Kerswell who obtained enlightening results
on statistical issues related to the dynamics of slugs and puffs for
the pipe flow transitional problem by modelling the flow through a
drastic reduction of the azimuthal resolution \cite{WK09}. In contrast
with them, we shall also probe the reliability of the approach at a
quantitative level by comparing results obtained when progressively
decreasing the resolution progressively.   

Gibson's well-known program \textsc{ChannelFlow}~\cite{CF} is used
throughout the present study, taking for granted that it has already
been abundantly validated and has served in many high resolution studies.
A first experiment, described in \S\ref{s2}, is focused on moderately
developed turbulence somewhat above $R_{\rm t}$. The second
experiment, in \S\ref{s3}, is devoted to a study of how the bifurcation
diagram is damaged by a reduction of the spatiotemporal resolution
in the transitional range: in a domain of size sufficient to contain
one wavelength of the band pattern, $R$ is decreased
by small steps from the turbulent regime down to the laminar state
and the values of $R_{\rm t}$ and $R_{\rm g}$ are systematically
recorded. From these two experiments, we determine an `optimal' resolution
above which the physics of the phenomenon is preserved, at the expense
of a tolerable shift of these two thresholds. Some preliminary results
supporting our prescription are presented in the last section ending
with a general discussion.

\section{Decreasing the numerical resolution:
qualitative effect on the turbulent state\label{s2}}

Fields and functions in our \textsf{C++} code are defined as in
\textsc{ChannelFlow}.  As to time integration, a backward formula
is taken, which treats the viscous term
implicitly and the nonlinear term explicitly. The time step
is adjusted so as to keep the CFL number below 0.4 while being maintained
smaller than 0.06~$h/U$ in all our simulations.
In the wall-normal direction, the spatial resolution is a function of the
number $N_y$ of Chebyshev polynomials used. The in-plane resolution depends
on the numbers $(N_x,N_z)$ of collocation points used in the
evaluation of the nonlinear terms. From the 3/2 rule applied to remove
aliasing, this corresponds to solutions evaluated in Fourier space
using $N'_{x,z}=2N_{x,z}/3$ modes, or equivalently to effective space
steps $\delta_{x,z}^{\rm eff}=L_{x,z}/N'_{x,z}=3L_{x,z}/2N_{x,z}$.
In the following, the
resolution will everywhere be specified using the triplet $(N_x,N_y,N_z)$.
In the range of Reynolds numbers of interest here (less than 500), simulations with $N_y\sim40$--$50$ and
$L_{x,z}/N_{x,z}$ less than $0.2$ are usually considered satisfactory.
Here, much lower resolutions have been considered in view of validating
our modelling attempt.

A first aim is at evaluating how well the fully turbulent state is
rendered when the spatial resolution is lowered. A systematic study has
thus been performed at moderate aspect ratio ($L_x=L_z=32$, lengths given
in units of the half-gap $h$, see note~\ref{fn1}) and $R=450$, a value at
which turbulence is permanently maintained. Initial conditions are taken
to be random and the
simulations rapidly reach a steady state regime, typically within less 
than 150 time units. Crude diagnostics about the flow regime are
obtained from an integral measure of the distance to the linear base flow derived from the quantity
$~\mathcal V^{-1}\!\int (u^2+v^2+w^2)\,{\rm d}x\,{\rm d}y\,{\rm d}z~$
built in {\sc ChannelFlow}, hereafter called $\Delta$.
($\mathcal V=2L_xL_z$ is the volume of the full three-dimensional domain.)
In order to get more local information, we have also considered
the space-time dependance of the velocity departure $u,v,w$ away from the
laminar base profile $u_{\rm b}=y\,\mathbf{\hat x}$, and especially of the
streamwise component $u$ in the plane $y=0$, $u_0\equiv u(x,y=0,z;t)$,
since it is a good tracer of any departure away
from the base state. As another
observable, we have considered the perturbation kinetic energy
$E_{\rm t}=\frac12(u^2+v^2+w^2)$, either at a point,
$E_{\rm t}(x,y,z;t)$, or its average over the gap,
$\bar E_{\rm t}(x,z;t)=\frac12\int_{-1}^{+1} E_{\rm t}(x,y,z;t)\,{\rm d}y$.
Fourier spectra of $u_0$ or $\bar E_{\rm t}$
have been extensively used.

\begin{figure*}
\begin{center}
\includegraphics[width=0.45\textwidth,height=0.34\textwidth,clip]%
{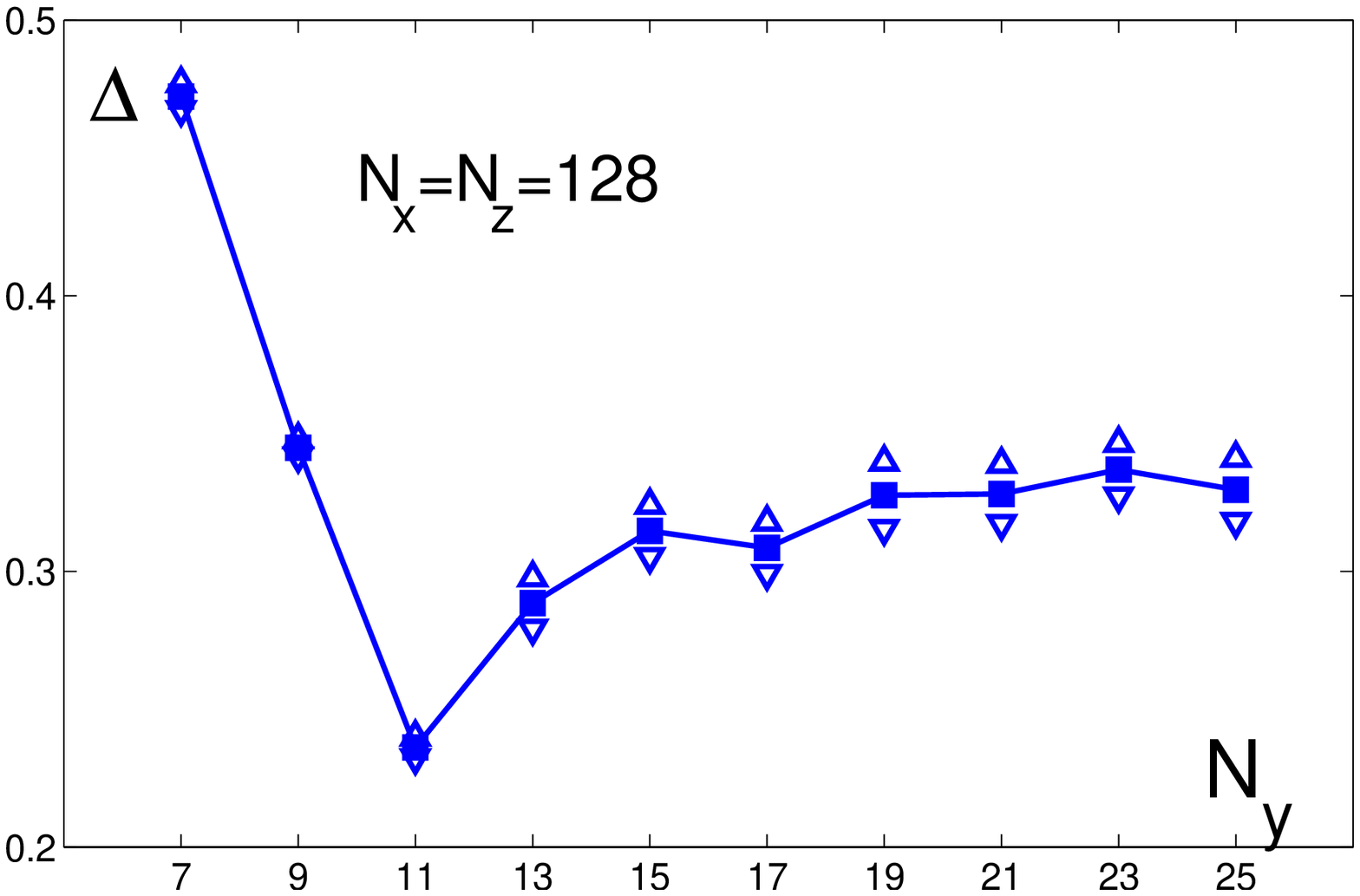}
\hspace{0.05\textwidth}
\includegraphics[width=0.45\textwidth,height=0.34\textwidth,clip]%
{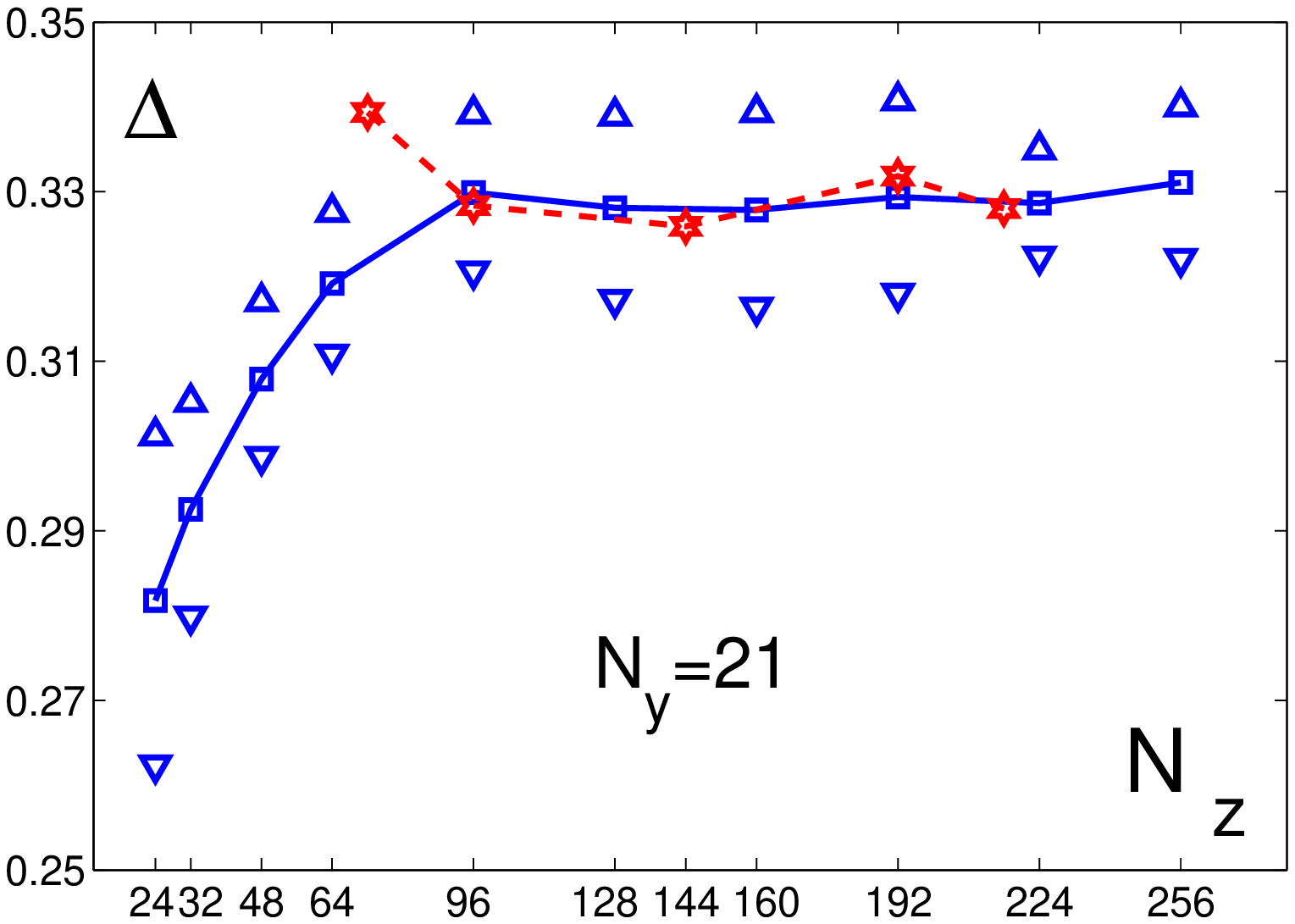}
\end{center}
\caption{Variation of time-averaged quantity $\Delta$ with the spatial
resolution of the pseudo-spectral scheme, as measured by the number of
modes $(N_x,N_y,N_z)$.
Left: variable number $N_y$ of Chebyshev polynomials. Right: variable
number of in-plane collocation points $N_x,N_z$.
With $L_x=L_z=32$, $R=450$. 
\label{f1}}
\end{figure*}

Figure~\ref{f1} displays the variation of the time average of $\Delta$ over the interval $t\in[400,2000]$ with $N_y$ (left) and
$N_x,N_z$ (right) for a system of size $L_x\times L_z=32\times32$
at $R=450$.
The experiment where $N_y$ is varied (left panel) is performed with
$N_x=N_z=128$.
Squares mark the time-averaged $\Delta$ and up/down triangles
indicate the standard deviation of its fluctuations around the average.
The experiment with variable $N_x,N_z$ (right panel) assumes $N_y=21$.
It reports two cases. The first one is with $N_x=N_z$ varying
from 24 to 256, where the squares indicate the time-averaged $\Delta$
and up/down triangles the standard deviation as before.
The second one is with $N_z=3N_x$ and the time-averaged $\Delta$
marked with stars, the corresponding standard deviation is the same
order of magnitude and not shown for the sake of readability.

\begin{figure*}
\begin{center}
\includegraphics[width=0.45\textwidth,clip]{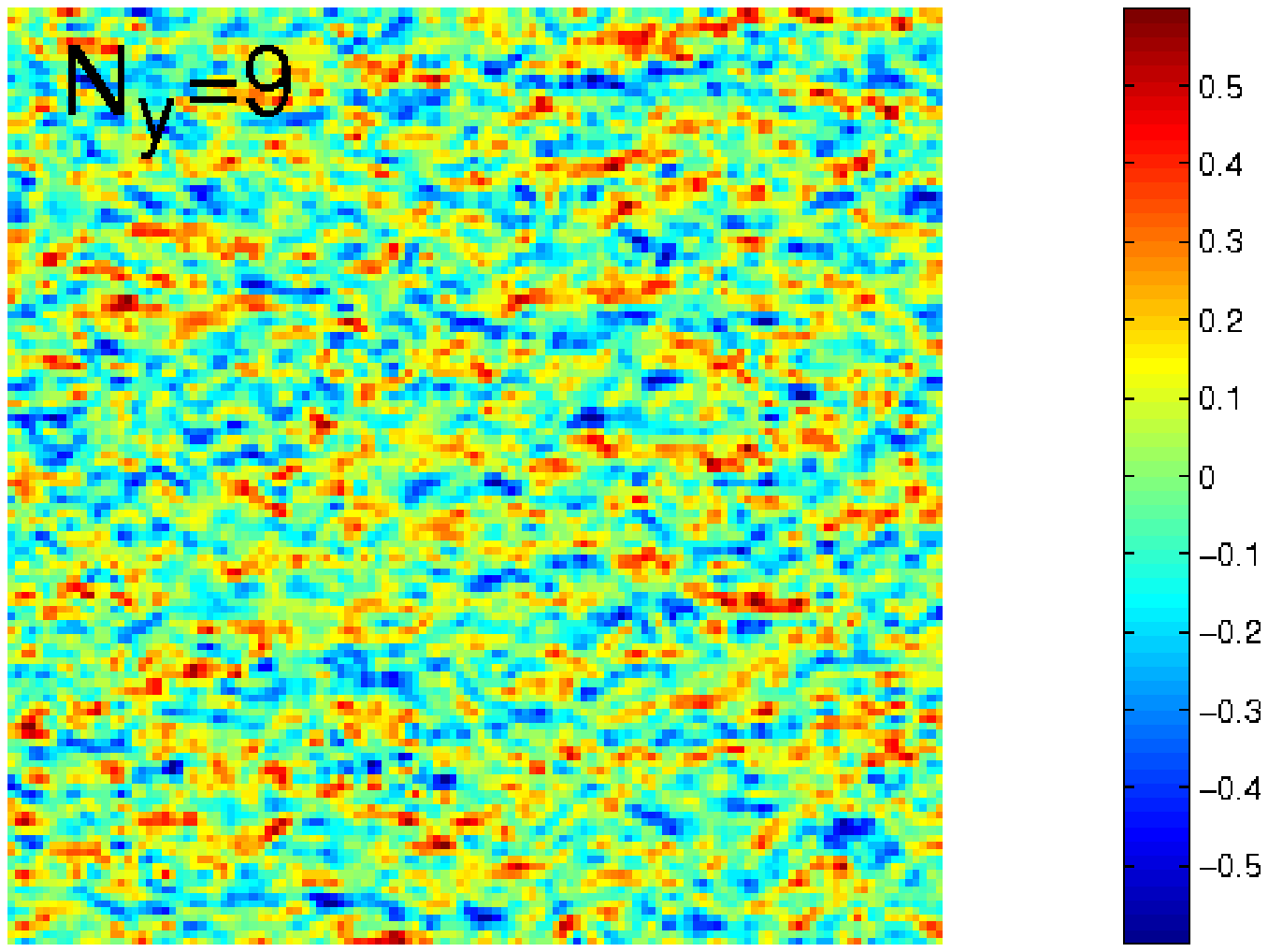}
\hspace{0.05\textwidth}
\includegraphics[width=0.45\textwidth,clip]{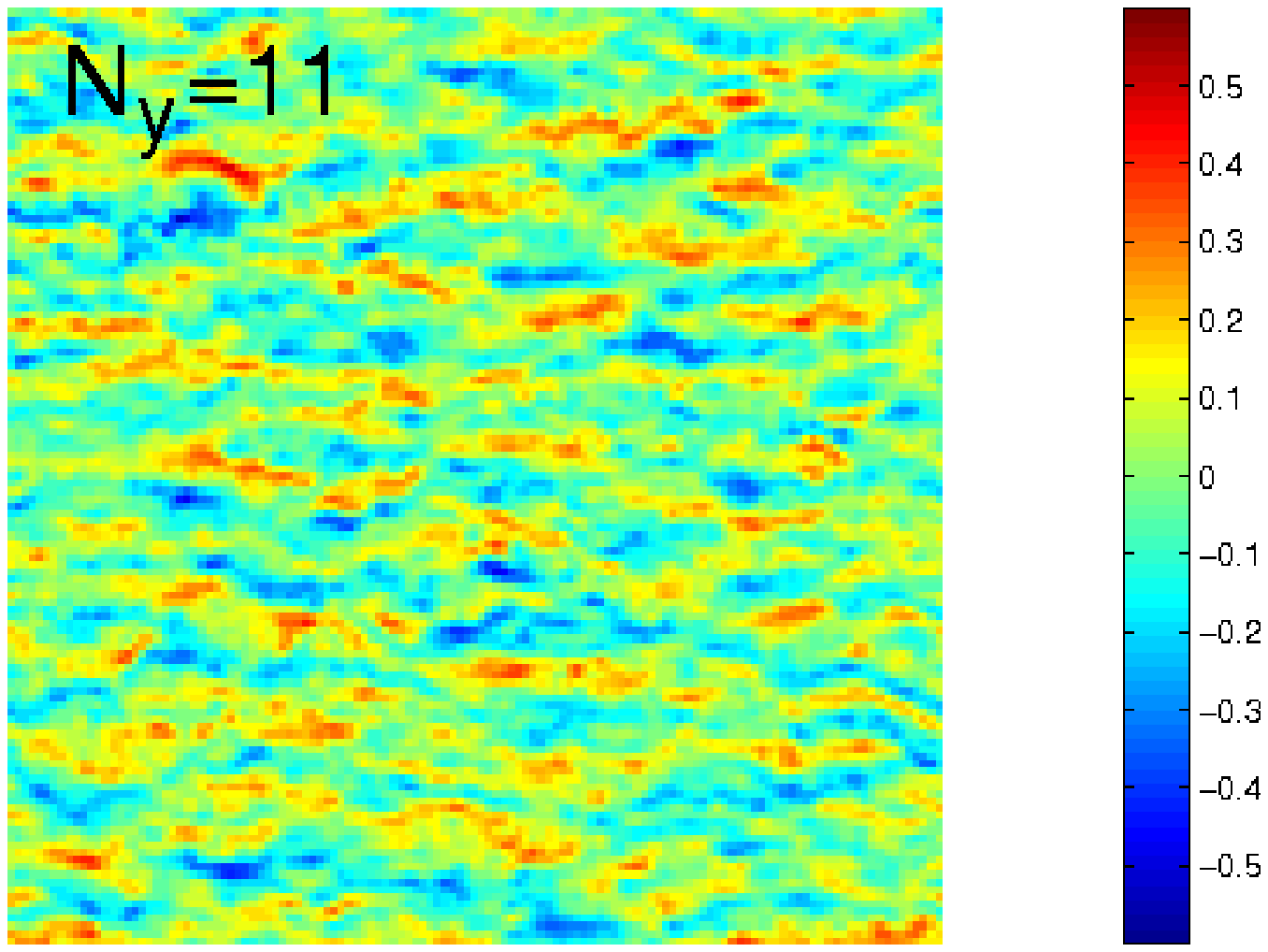}
\vspace{2ex}

\includegraphics[width=0.45\textwidth,clip]{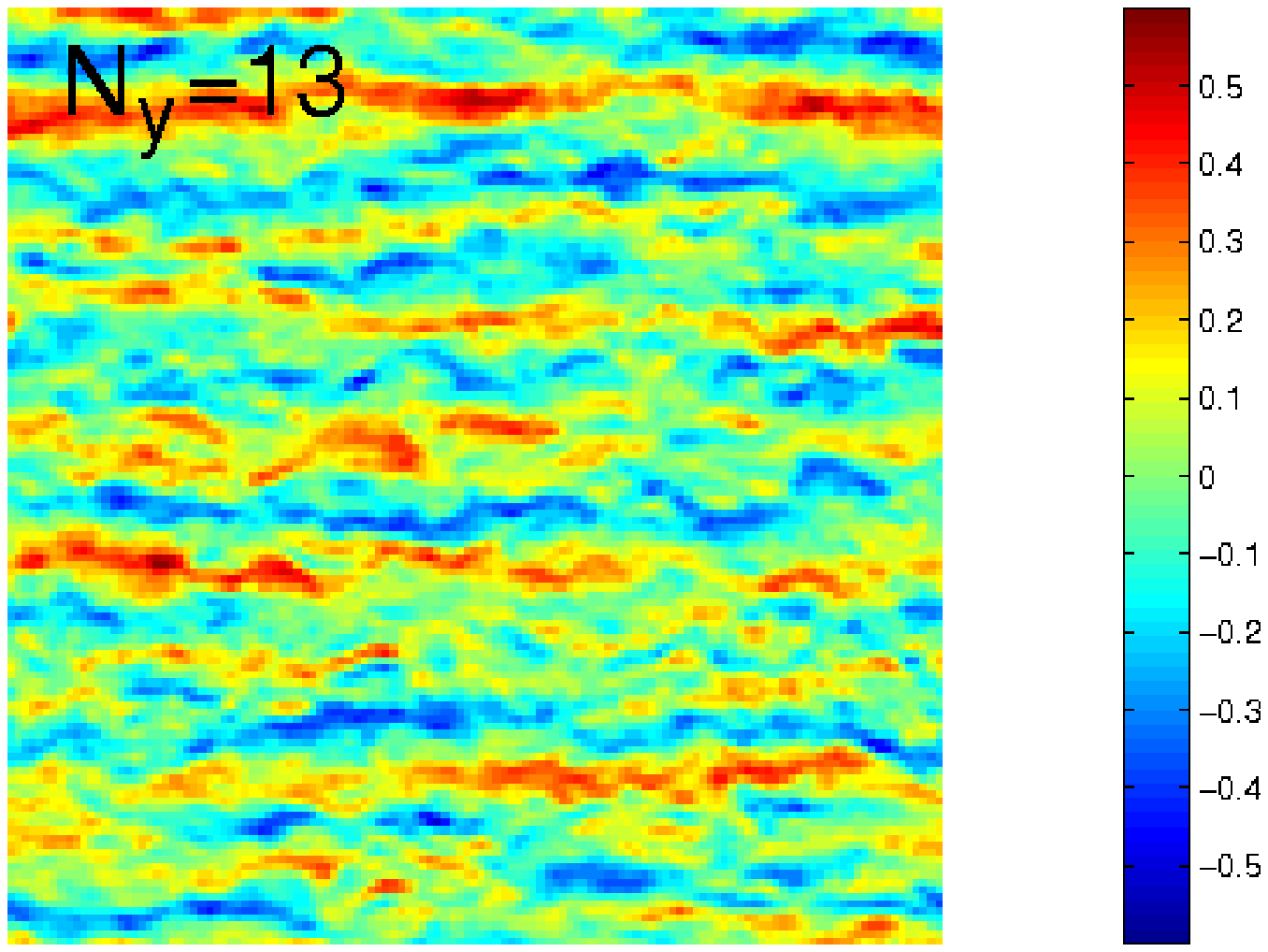}
\hspace{0.05\textwidth}
\includegraphics[width=0.45\textwidth,clip]{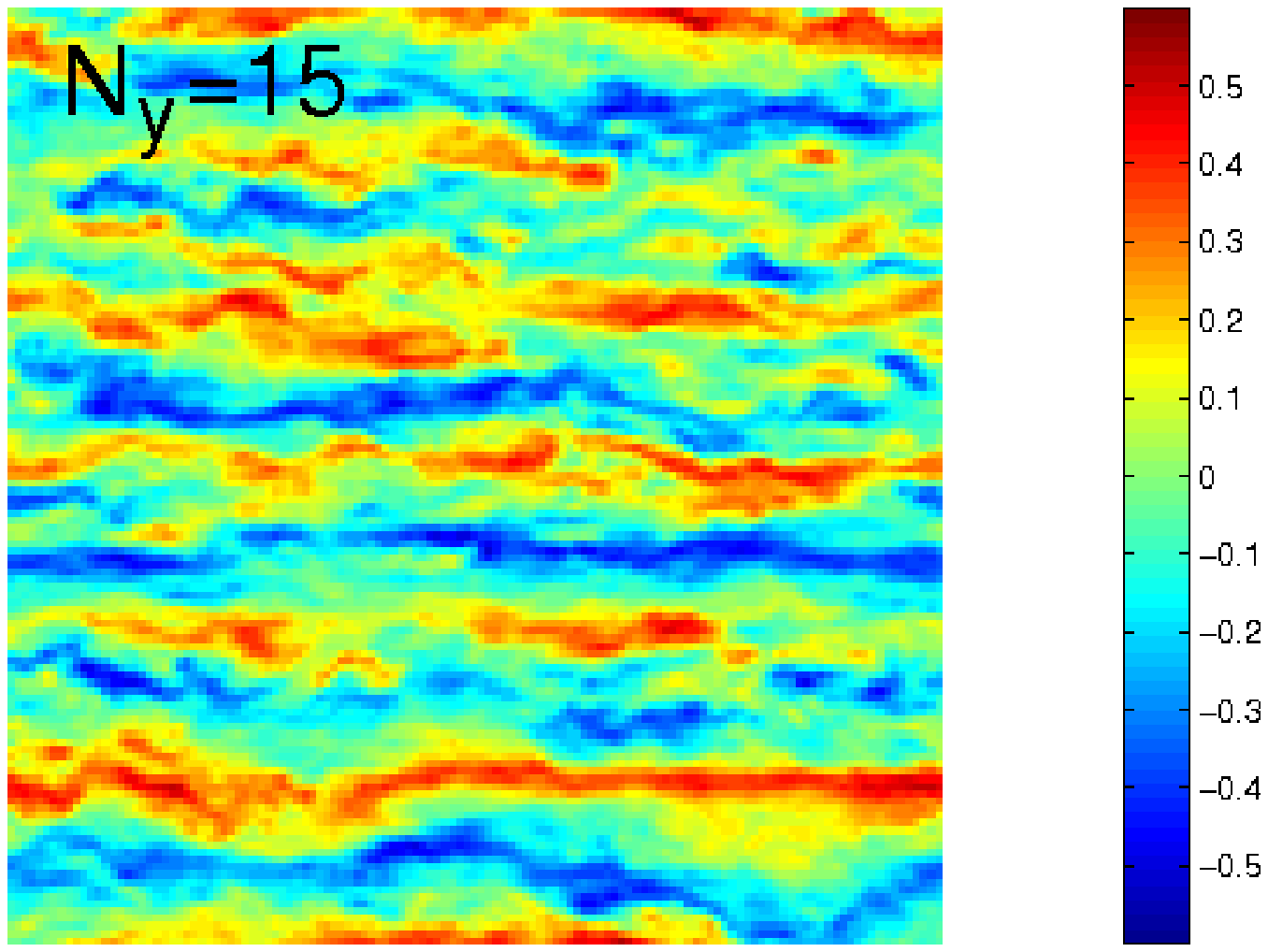}
\vspace{2ex}

\includegraphics[width=0.45\textwidth,clip]{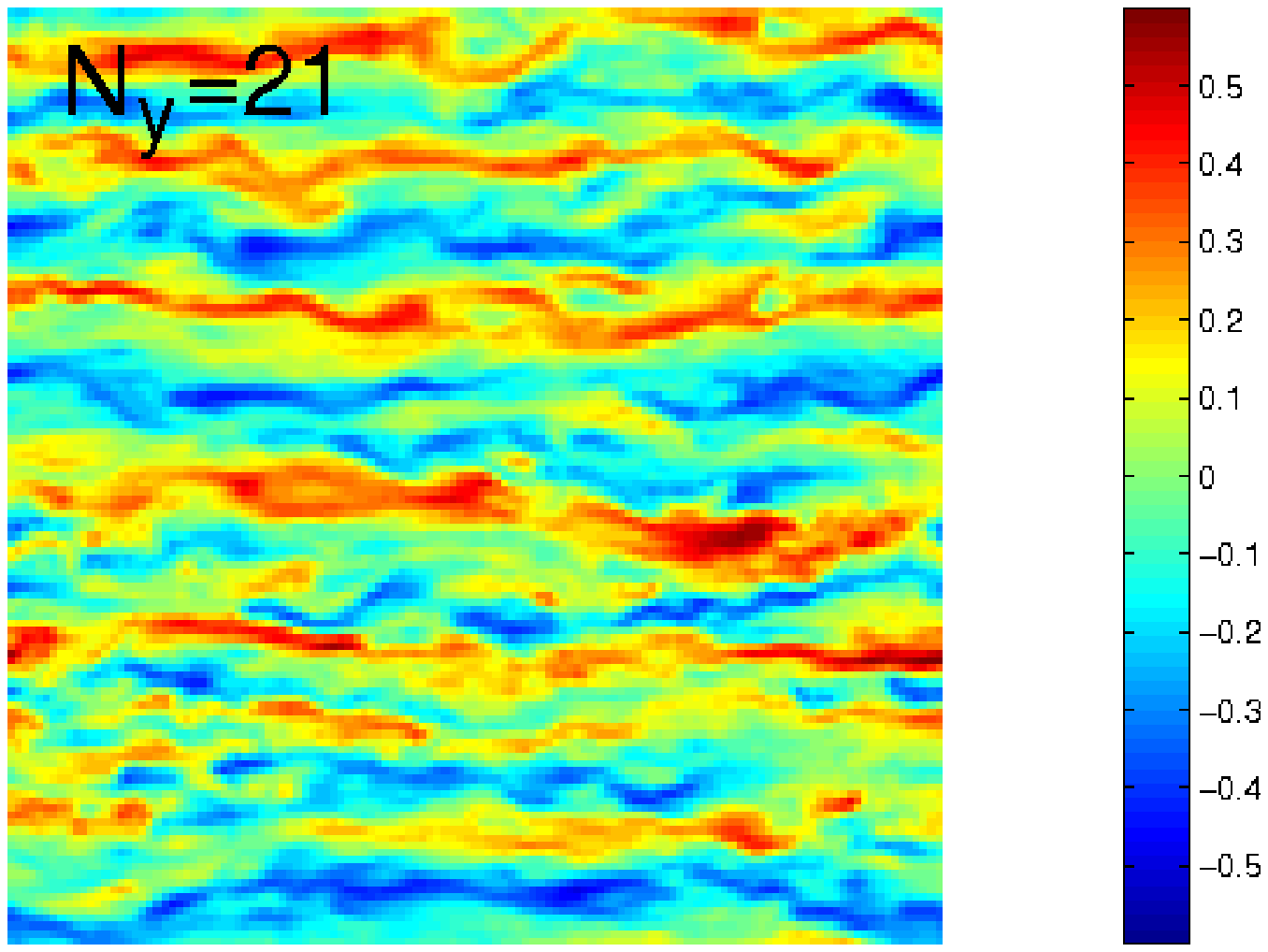}
\hspace{0.05\textwidth}
\includegraphics[width=0.45\textwidth,clip]{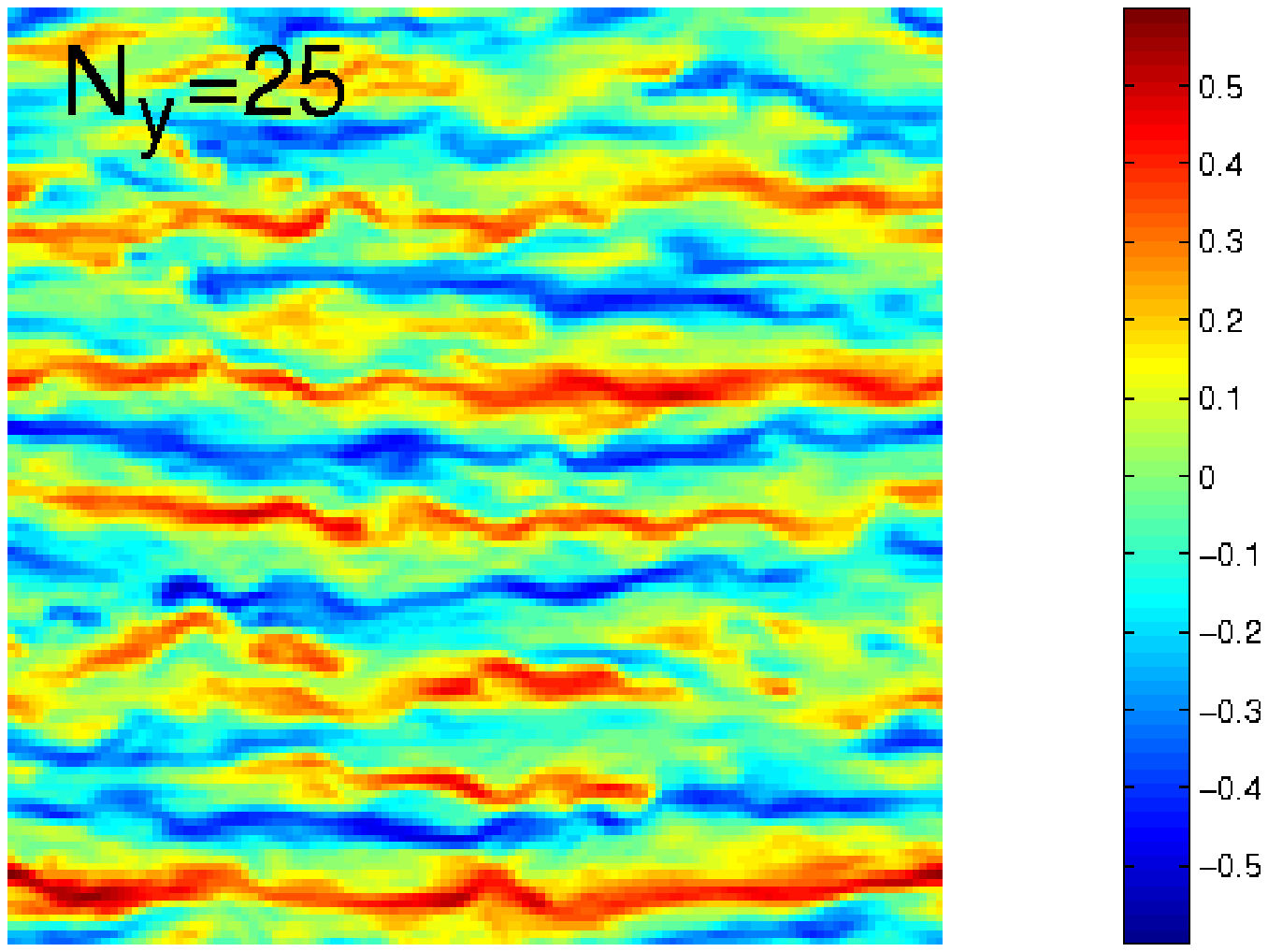}
\end{center}
\caption{Colour level representations of the $x$ component of the
velocity perturbation in the mid-plane $y=0$ of full three dimensional
snapshots of the numerical solutions obtained in a domain
$32\times2\times32$ for $R=450$ at $t=2000$.
$N_x=N_z=128$ in-plane collocation points, and $N_y$ Chebyshev
polynomials in the cross-stream direction, with $N_y$ ranging from 9 to 25. 
The $x$ and $z$ axes are respectively horizontal and vertical, like in all most of the other figures displaying patterns, except stated otherwise.}
\label{f2}
\end{figure*}

A clear convergence of the results is observed as the resolution is
increased, marked by a plateau reached for $N_y\sim 15$ and
$N_x=N_z\sim96$. We are rather interested in the opposite limit
of decreased resolution. Whereas a gentle trend is observed for
$N_y\ge11$, the cases $N_y=9$ and $7$ behave differently and a divergence
in finite time is even observed for $N_y<7$. In fact, for the moderate
Reynolds numbers of interest here, $N_y=11$ appears to be a bound below
which the numerical solution is no longer physical. Figure~\ref{f2}
displays the streamwise component of the disturbance in the plane $y=0$.
For $N_y=9$ the solution shows an
anomalously fine in-plane structure, as if turbulent energy could only
be dissipated by being transferred to structures with fast variations
in $x$ and $z$. By contrast, the solutions obtained for $N_y\ge11$
display coarser velocity fluctuations, the case $N_y=13$ being already
qualitatively similar to those for $N_y=15$, $21$, and $25$. These
patterns are reminiscent of the very large streamwise streaky turbulent
structures obtained in wall-bounded flows \cite{Ketal96,ptrsa07}.

\begin{figure*}
\begin{center}
\includegraphics[width=0.45\textwidth,clip]{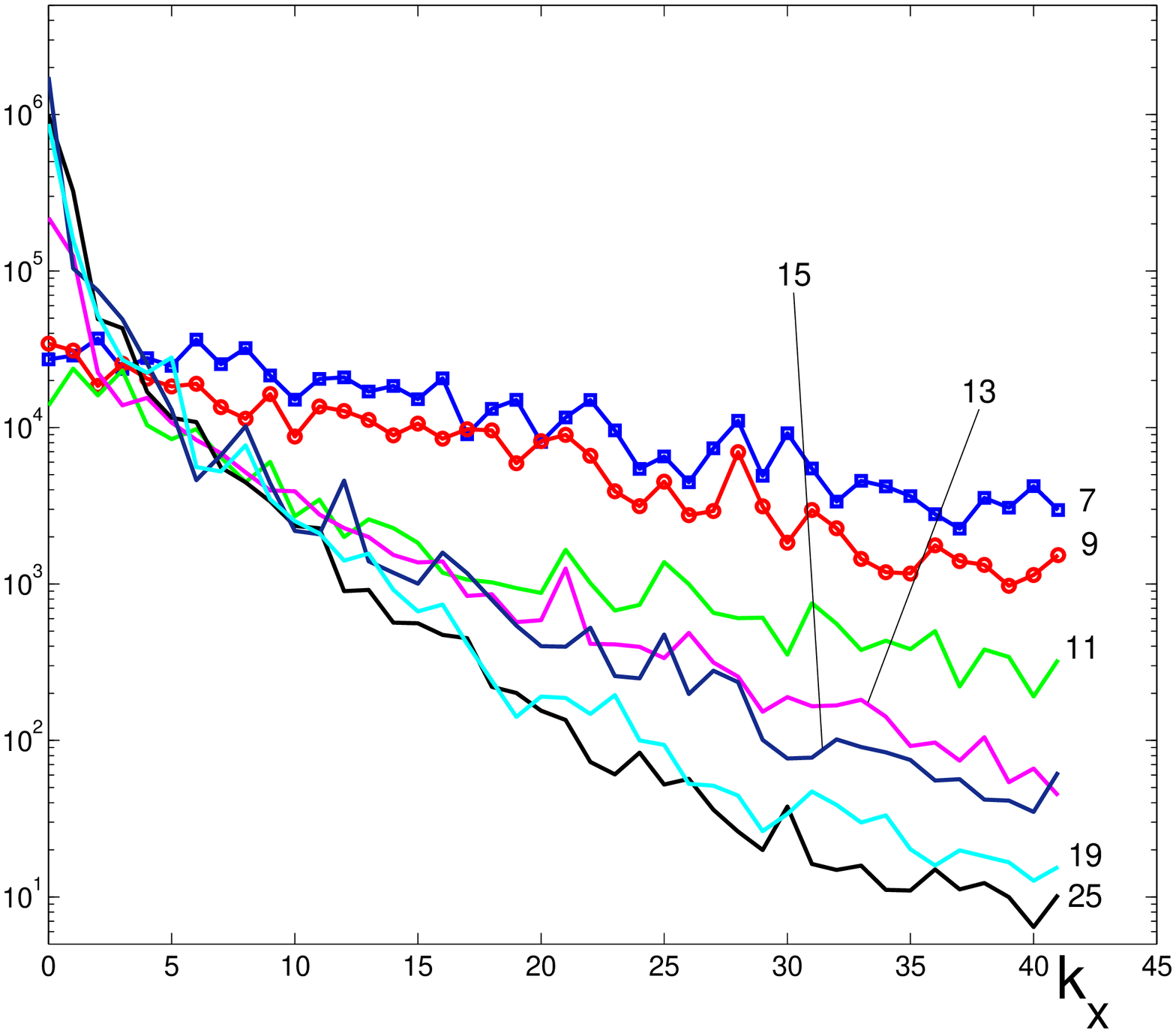}
\hspace{0.05\textwidth}
\includegraphics[width=0.45\textwidth,clip]{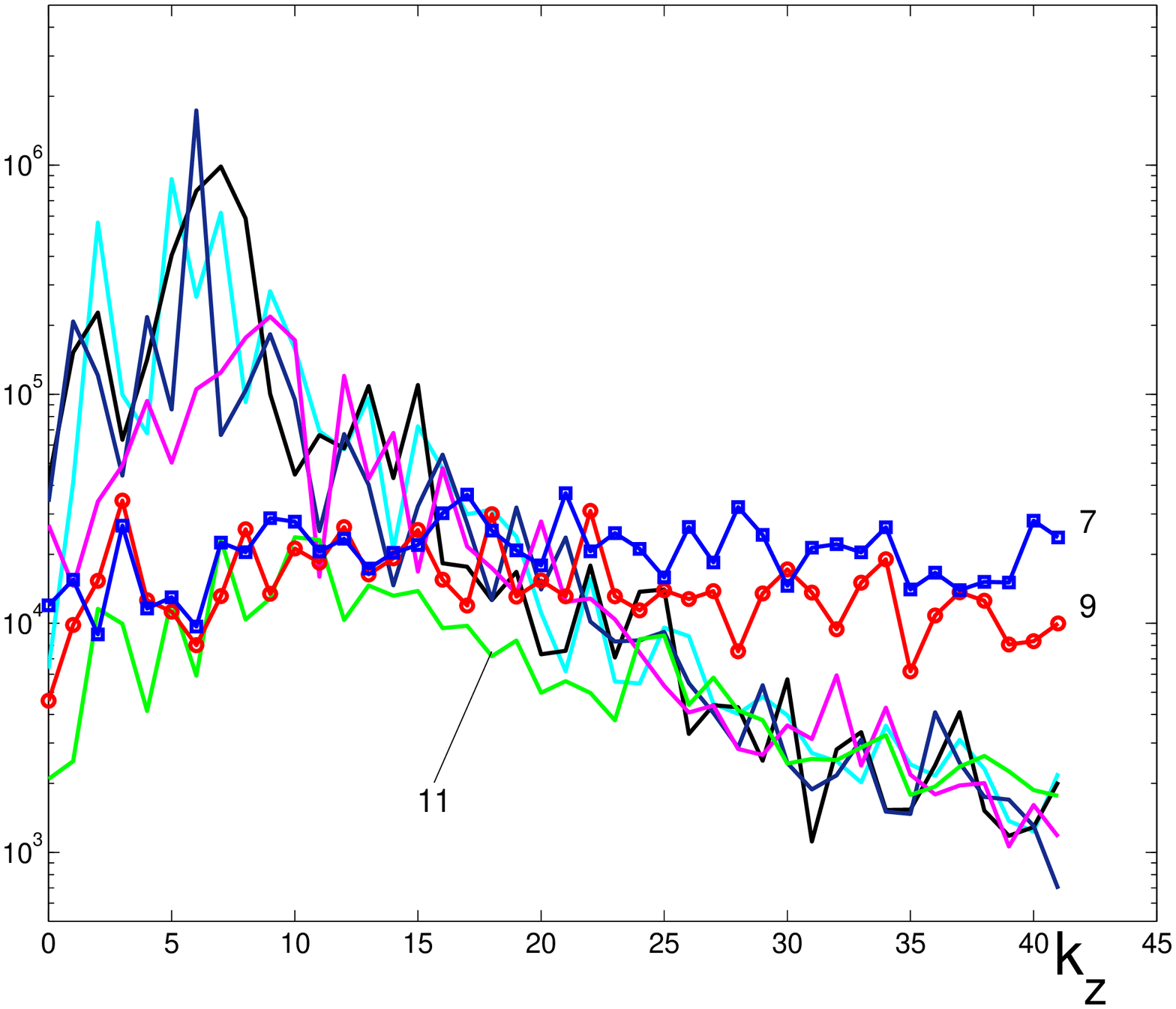}

\includegraphics[width=0.45\textwidth,clip]{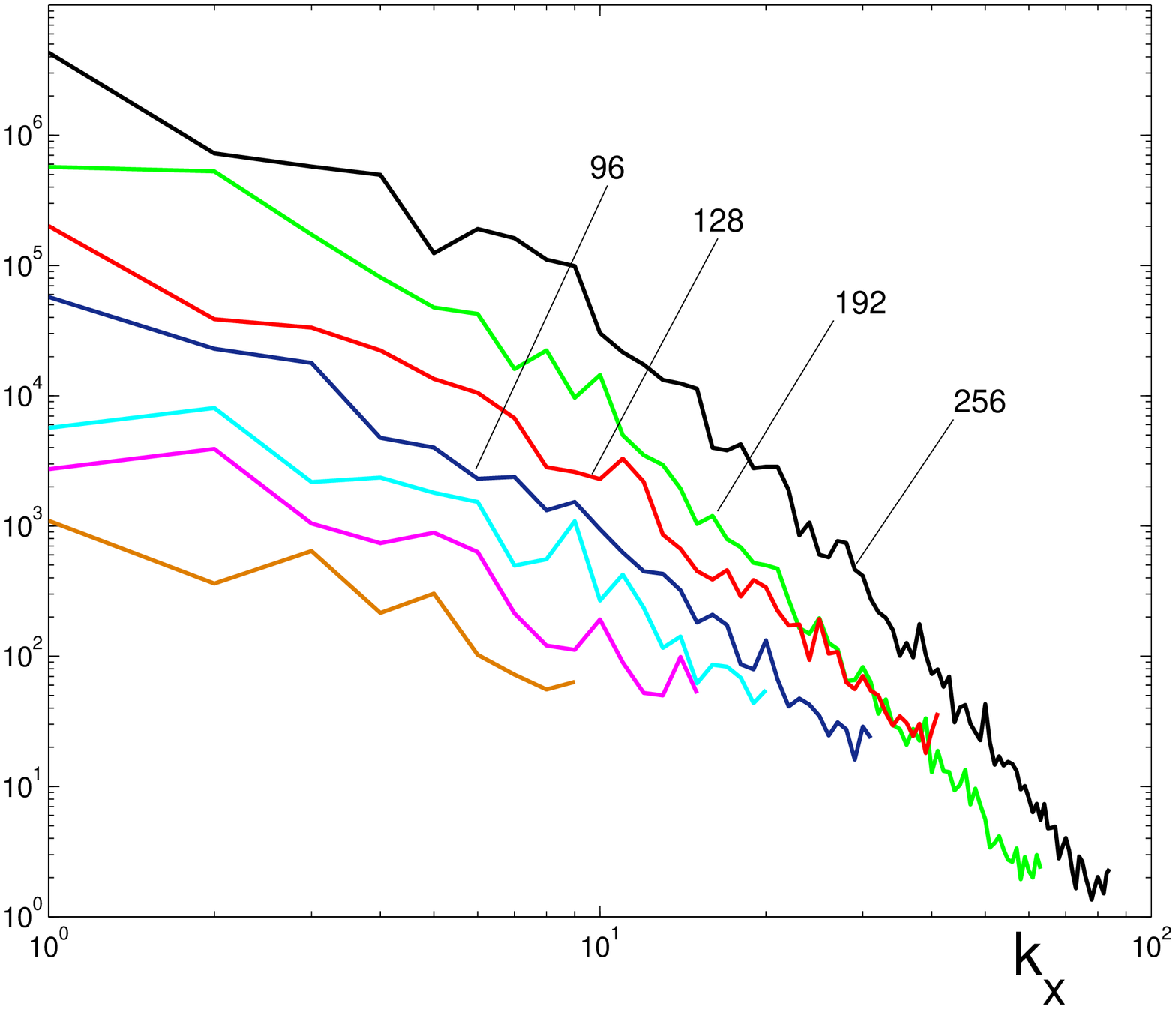}
\hspace{0.05\textwidth}
\includegraphics[width=0.45\textwidth,clip]{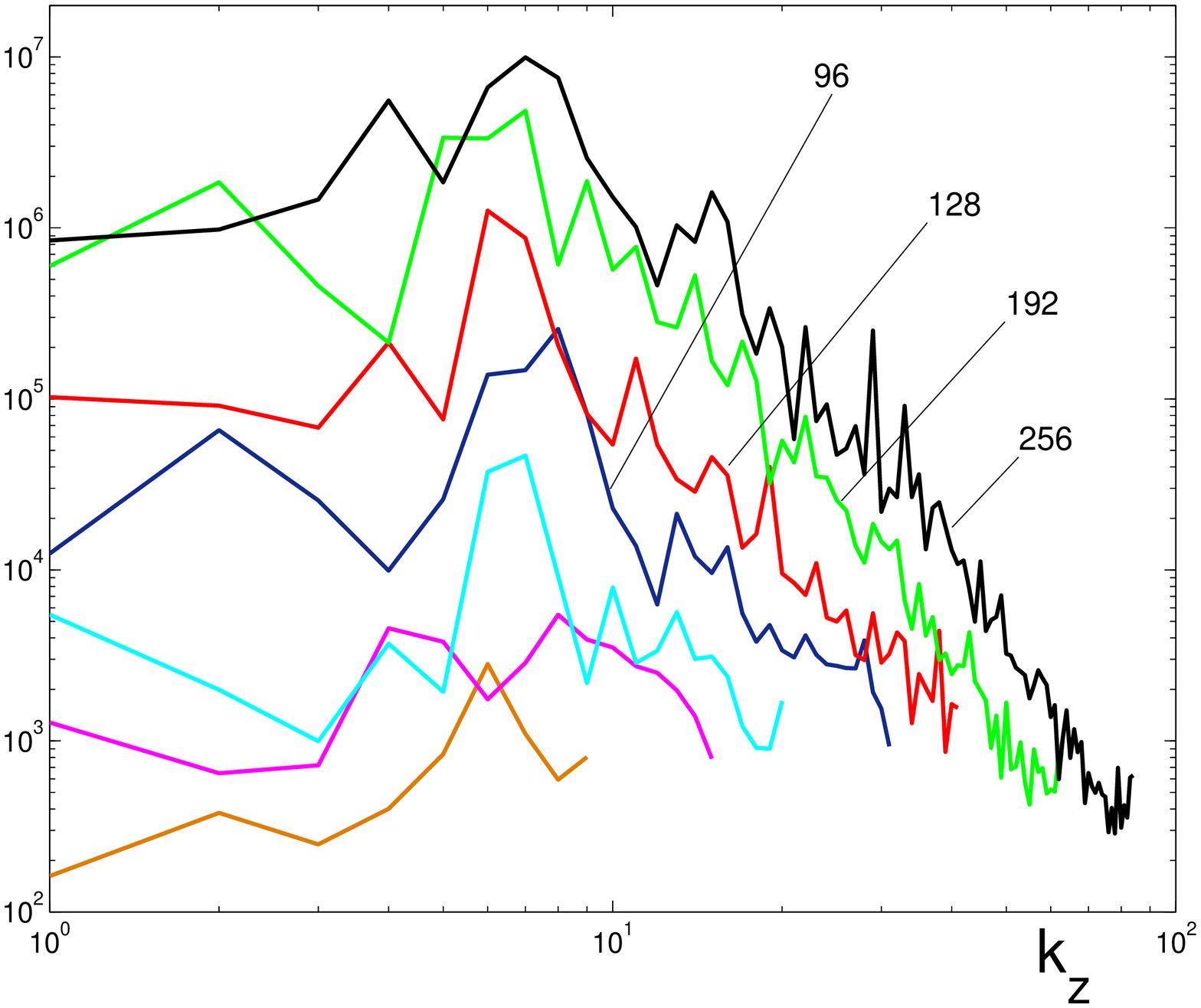}
\end{center}
\caption{Envelopes of Fourier spectra of the streamwise velocity component
in the mid-plane $u_0$ for $L_x=L_z=32$, $R=450$, and a single
snapshot at $t=2000$. Top: $N_y$ variable, $N_x=N_z=128$.
Bottom: $N_x=N_z$ variable, $N_y=21$. Spectra are not scaled by the total
number of modes $N_xN_z$. For the top line $N_xN_z$ is constant and
spectra envelopes appear on top of each other, whereas for the bottom
line the number of modes depend on the resolution so that they are
shifted upward as it increases.\label{f3}}
\end{figure*}

These structures can further be identified in Figure~\ref{f3}
which presents Fourier spectra of $u_0(t=2000)$ for the different
resolutions considered. In all the graphs, the envelopes of the projections
of the spectra are displayed, that is to say 
$\Sigma(k_x)=\max_{k_z} S(k_x,k_z)$ and
$\Sigma(k_z)=\max_{k_x} S(k_x,k_z)$,
where $S(k_x,k_z)=|\hat u_0|^2$ is the Fourier spectrum of
$u_0$. The information contained in these quantities, though somehow limited,
is more readable than the display of the full spectra while giving a proper
account of the anisotropy of the flow that clearly distinguishes
the spanwise and streamwise directions.

For variable $N_y$, Fig.~\ref{f3} (top), the cases
$N_y=7$ and $9$ are clearly not properly resolved since
the envelopes of the projected spectra do not decay as $k_x$ and $k_z$ increase.%
\footnote{The units for $k_{x,z}$ are $2\pi/L_{x,z}$. The maximum value
of $k_{x,z}$ is $N_{x,z}/3$ because the solution is represented
using $N'_{x,z}=2N_{x,z}/3$ and the spectrum goes from $-N'_{x,z}/2$ to
$N'_{x,z}/2$ with complex conjugation relations.} Already present for $N_y=11$, the
correct tendency strengthens as $N_y$ increases, though less rapidly in the
streamwise direction than in the spanwise direction. Along the $k_z$ axis,
except for $N_y=11$, a clear peak is observable for $k_z\sim7$, i.e.
$\lambda_z=L_z/7)\sim4.6$; this value is somewhat larger than but of the
same order of magnitude as the width $\ell_z$ of the MFU mentioned above.
In contrast, no peak at finite $k_x$ is observed in the streamwise
direction but a monotonic decrease as $k_x$ varies from zero to
$k_{x,{\rm max}}=N_x/3$.
These two features are in line with the observation of the very elongated
streamwise streaky structures in wall-bounded flows.
Similar trends are observed for $N_y=21$ and variable $N_x$ in
Figure~\ref{f3} (bottom) where the spectra appear shifted with respect
to one another due to the absence of normalisation by the number of modes
in order to improve the readability of the figure. The peak at $k_z\sim7$
is visible for $N_{x,z}\ge64$ but not at lower resolution which is not
surprising when we compare the corresponding wavelength $L_z/7\approx4.6$
and the effective space step $\delta_{x,z}^{\rm eff}$: for $N_{x,z}=48$,
this makes $\delta_{x,z}^{\rm eff}=3L_{x,z}/2N_{x,z}=1$, hence less than
five points per spanwise wavelength, which really does not seem to be enough.

\begin{figure*}
\begin{center}
\includegraphics[width=0.45\textwidth,clip]{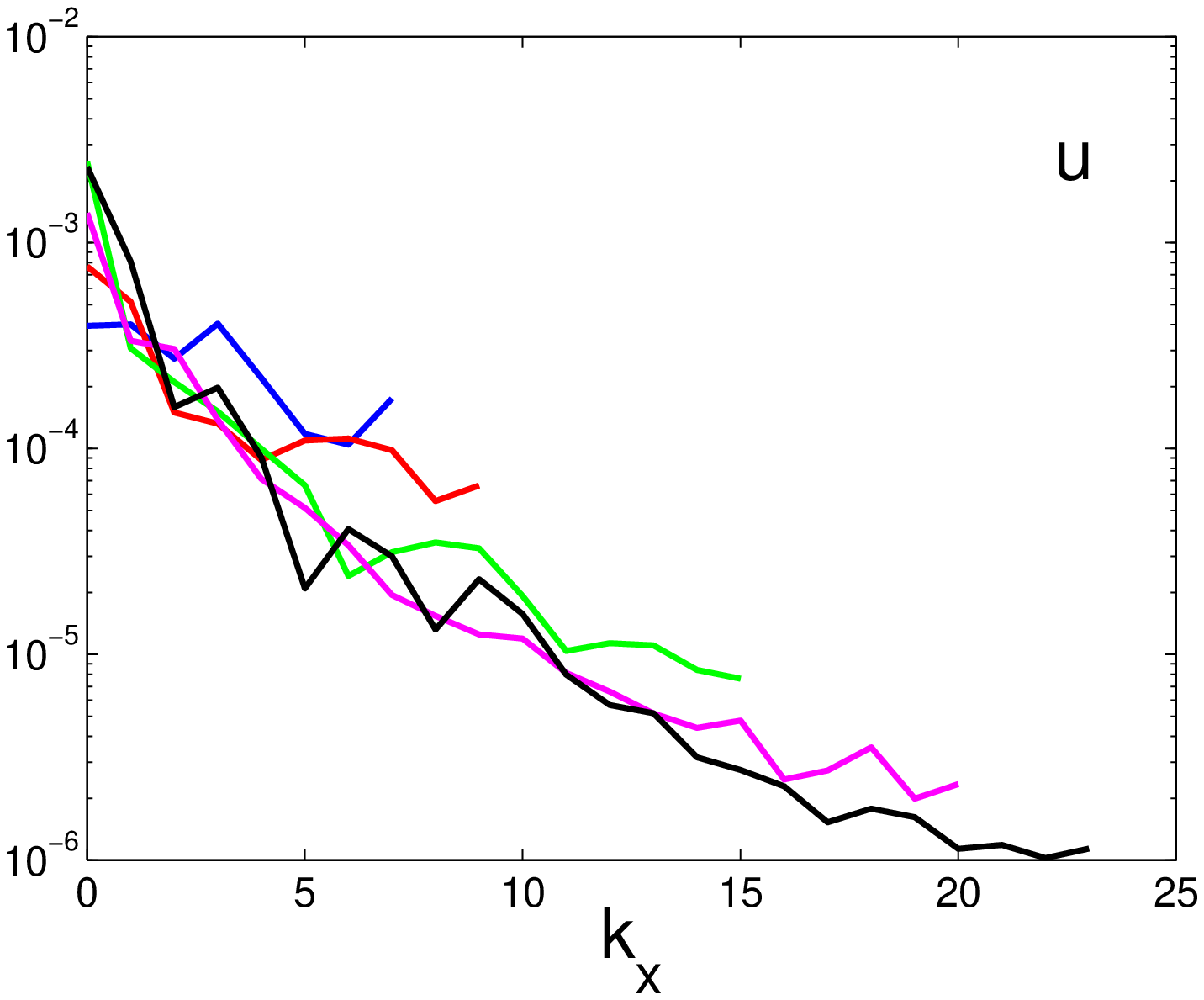}
\hspace{0.05\textwidth}
\includegraphics[width=0.45\textwidth,clip]{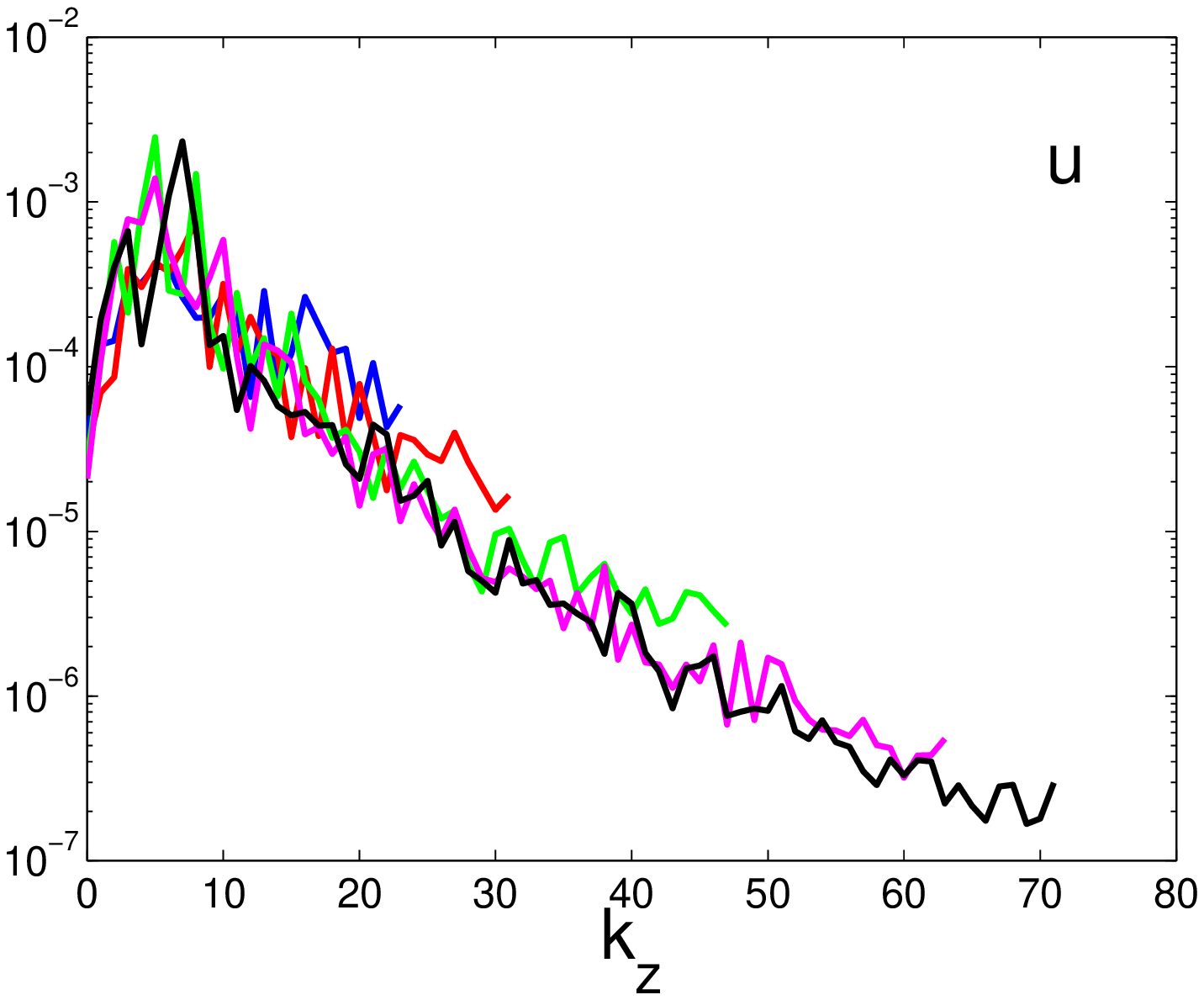}

\includegraphics[width=0.45\textwidth,clip]{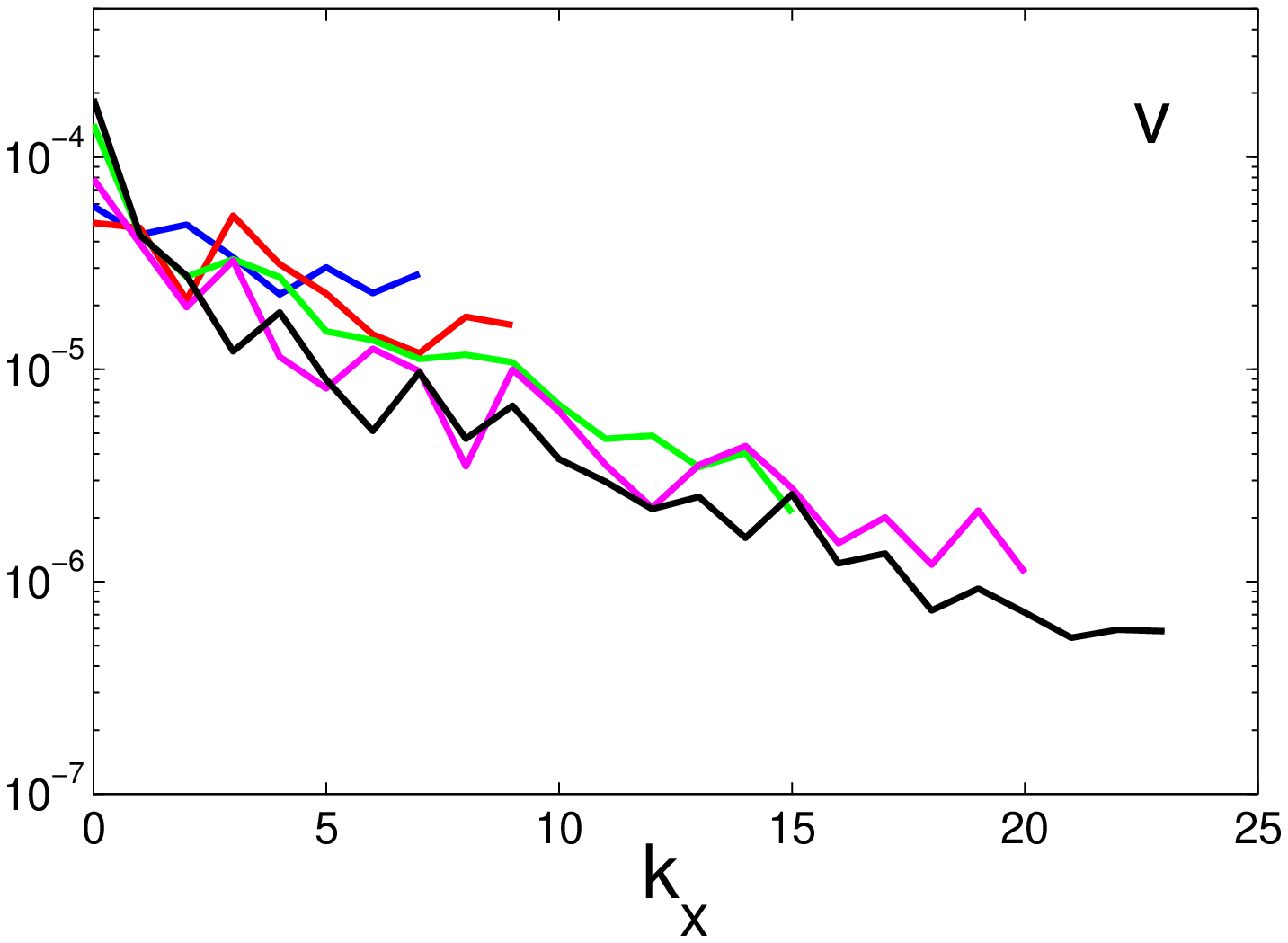}
\hspace{0.05\textwidth}
\includegraphics[width=0.45\textwidth,clip]{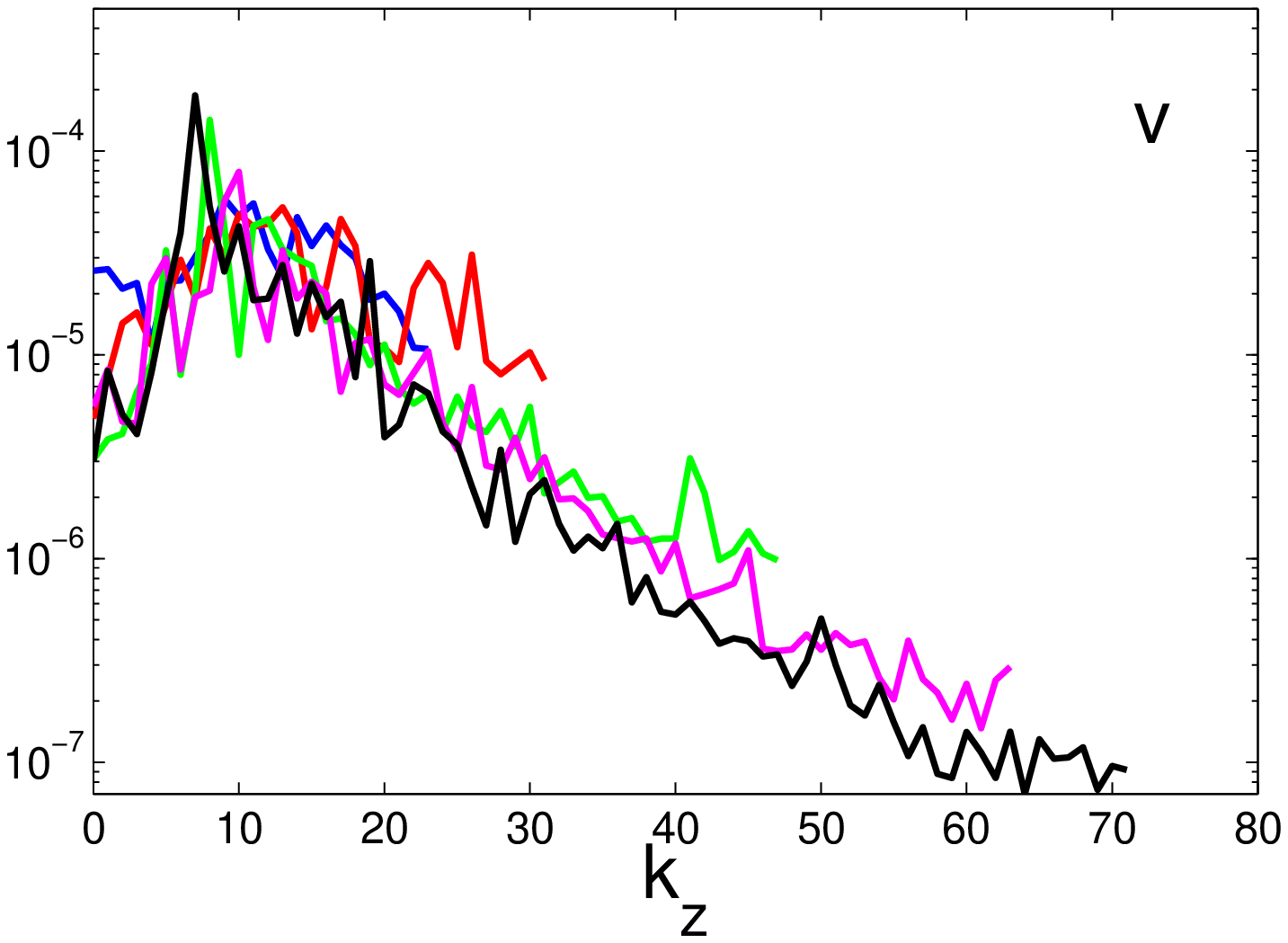}

\end{center}
\caption{\label{f4}Fourier spectra of the streamwise ($u$) and wall-normal
($v$) corrections to the laminar profile for the turbulent flow at $R=450$,
evaluated at $y=0$ ($L_x=L_z=32$, single
snapshot at $t=2000$, $N_y=21$ and $N_z=3N_x$ variable).
Top: streamwise perturbation $u$. Bottom: wall-normal perturbation $v$.
Left: streamwise component $k_x$. Right: spanwise component $k_z$.
Successively: $(N_x,N_z)=$ $(24,72)$, $(32,96)$, $(48,144)$, $(64,192)$,
and $(72,216)$. All spectra rescaled by $N_xN_z$ in order to show how
well they pile up.}
\end{figure*}

Up to now we have not yet taken into account the fact that the velocity
fluctuations vary much more rapidly in the spanwise direction than
in the streamwise direction. Here we thus consider effective space steps
systematically smaller by a factor of three along $z$ than along $x$, i.e.,
$N_x=N_z/3$. Results for $\Delta$ are shown in the right panel of
Figure~\ref{f1} as stars. Except for the lowest value $(N_x,N_z)=(24,72)$,
all other values agree with those determined for $N_x=N_z$ when plotted as
functions of $N_z$, which means that in-plane resolution can be appreciated
from the value of $N_z$ alone with $N_x$ down to a factor of three smaller.
This observation is confirmed by the examination of the envelope spectra as
defined above shown in figure~\ref{f4}.
Both $u_0$ and $v_0\equiv v(x,y=0,z;t)$ at $t=2000$ have been considered.
Here the Fourier amplitudes have been normalised by the corresponding
numbers of modes and it is clearly seen that they pile up for $N_x\ge48$,
$N_z\ge144$, that is to say   $\delta_x^{\rm eff}=3L_x/2N_x=1$ and
$\delta_z^{\rm eff}=3L_z/2N_z=0.33$.

\begin{figure*}
\begin{center}
\includegraphics[height=0.32\textwidth,width=0.45\textwidth,clip]%
{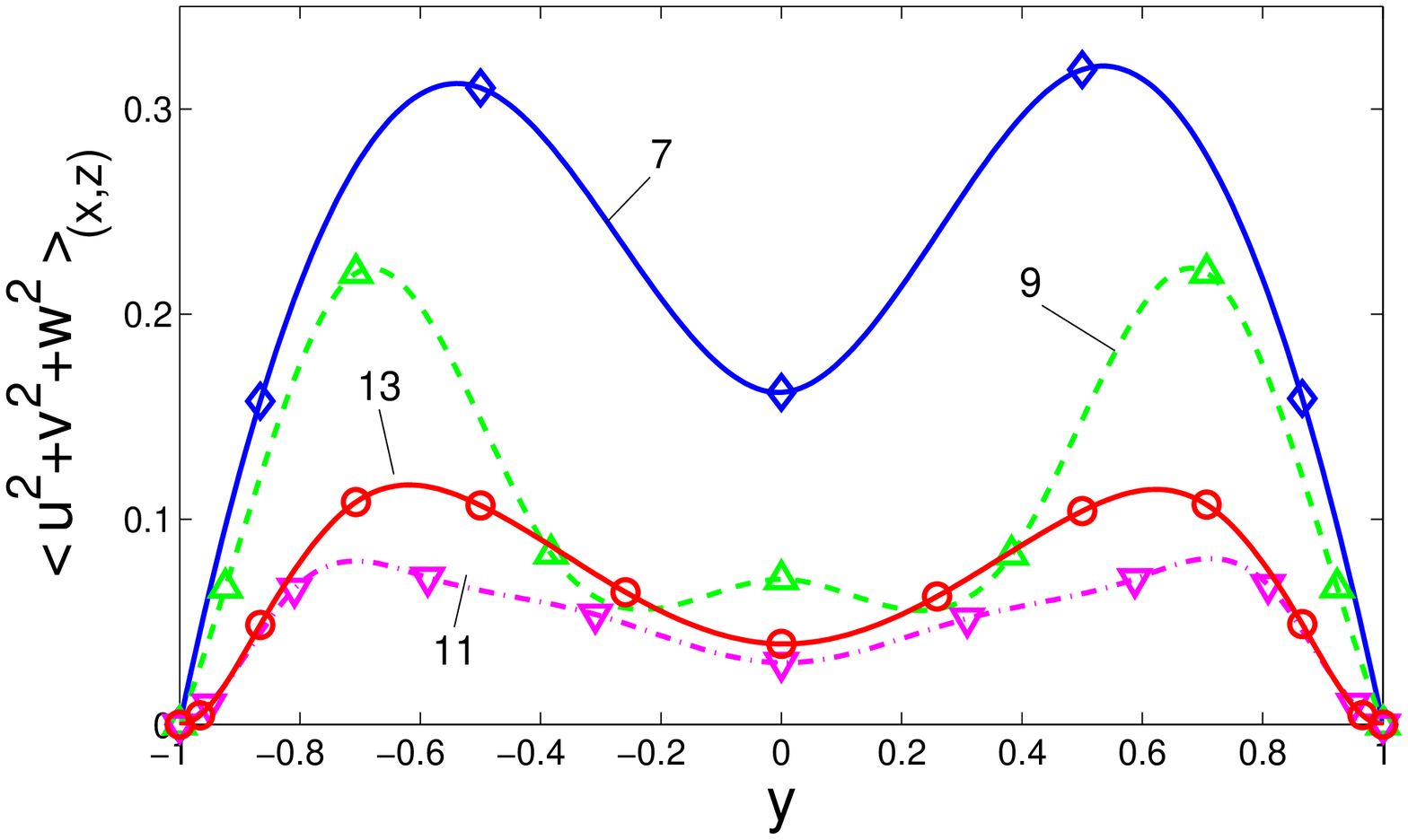}
\hspace{0.05\textwidth}
\includegraphics[height=0.32\textwidth,width=0.45\textwidth,clip]{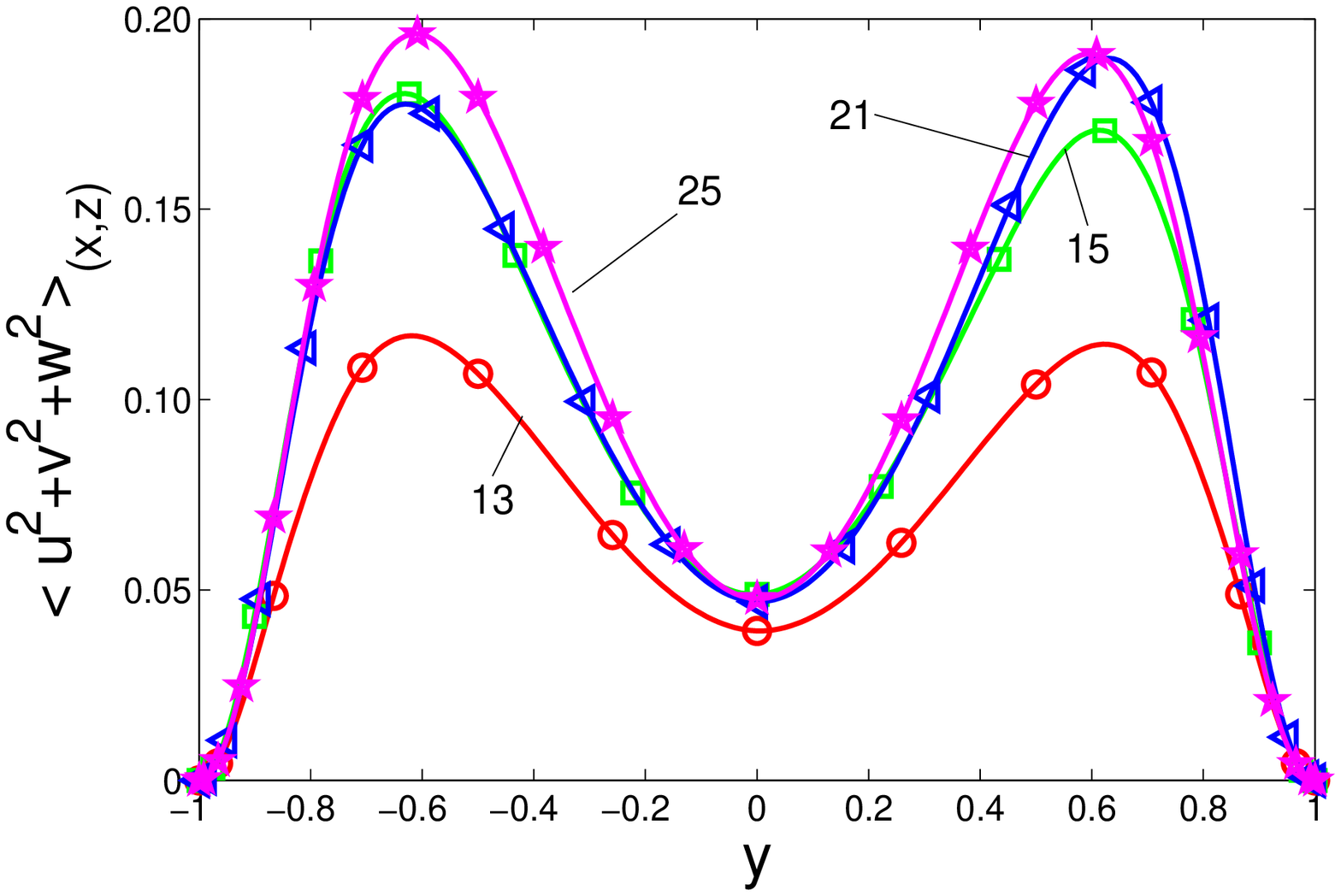}
\end{center}
\caption{Profiles of $\bar\Delta(y)$ as functions of the wall-normal
resolution $N_y$ as interpolated from values at collocation points using
cubic splines. Left: lowest resolution, $N_y$ from 7 to 13. Right: highest
resolutions, $N_y$ from 13 to 25. ($L_x=L_z=32$,
$N_x=N_z=128$, $R=450$, $t=2000$.)\label{f5}}
\end{figure*}

How well the turbulent regime is reproduced in low resolution simulations
has also been tested by comparing in-plane averaged profiles of the streamwise
velocity correction $\bar u(y)=(1/L_xL_z)\int u(x,y,z)\,{\rm d}x\,{\rm d}z$
and the squared distance to the base solution $\bar \Delta(y)=
(1/L_xL_z)\int (u^2+v^2+w^2)\,{\rm d}x\,{\rm d}z$. Figure~\ref{f5} displays
the latter quantity for a series of experiments with $N_y$ varying from
7 to 25. Results for $\bar u(y)$ are similar. No averaging over time has
been performed: a single snapshot at $t=2000$ has been used in each case,
which explains possible irregularities and a slight lack of $y$-symmetry
for $N_y=15$ and 21. From the figure, one can immediately see that
profiles for $N_y=7$ and 9 are off, that those for $N_y=11$ and $13$
progressively evolve so as to tend toward a limiting profile, and that
consistent results are obtained for $N_y\ge15$ in agreement
with observations already made when considering Figure~\ref{f1} (left).

\section{Decreasing the numerical resolution: quantitative effect on the
bifurcation diagram\label{s3}}

This preliminary study suggests that the featureless turbulent regime
is reasonably well rendered provided that $N_y\ge15$ and
effective in-plane resolution $\delta_x^{\rm eff}=3L_x/2N_x\le1$ and
$\delta_z^{\rm eff}=3L_z/2N_z\le0.33$. It remains to test the effect
of a lowering of the resolution on the peculiarities of the
`turbulent$\to$laminar' transition {\it via\/} oblique turbulent bands,
which is done in the present section.

According to the experiments \cite{Petal03}, the pattern can be
characterised by a wavevector $\mathbf{k}^{\rm patt}$ with components
$k_x^{\rm patt}=2\pi/\lambda_x$ and $k_z^{\rm patt}=2\pi/\lambda_z$.
Furthermore, $\lambda_x\simeq110$ is obtained all along the transitional
range and $\lambda_z$ varies from $50$ for $R=395$ close to the top at 
$R_{\rm t}\simeq410$, to $82$ at $R=340$ when the bands break down into
patches that finally decay below $R_{\rm g}\simeq325$. Wishing to perform
simulations in wide domains at the lowest possible computational cost, we
now validate our modelling strategy by examining the effect of a lowering
of the resolution on the bifurcation diagram in extended domains of size
sufficient to fit at least about one streamwise wavelength $\lambda_x$
and one spanwise wavelength $\lambda_z$ of the band pattern. Accordingly,
we consider rectangular domains of size $L_x\times L_z$ with $L_x\ge110$
and $L_z\ge50$. Except for $N_y=7$ and $9$, oblique bands are obtained
without any difficulty in a full range of Reynolds numbers.

Experiments are all done under the same protocol: a featureless turbulent
regime is prepared at $R=450$ and $R$ is next decreased by steps. At every
value of $R$, the simulation is pursued sufficiently long for the
establishment of statistical equilibrium before further decreasing $R$.
Usually, this takes at least 5000 time units. The distance $\Delta$ is
recorded as a function of time. After elimination of a transient just
after the change in $R$, its time average and the rms value of its fluctuations are computed.

\begin{figure}
\includegraphics[width=0.45\textwidth,height=0.34\textwidth,clip]%
{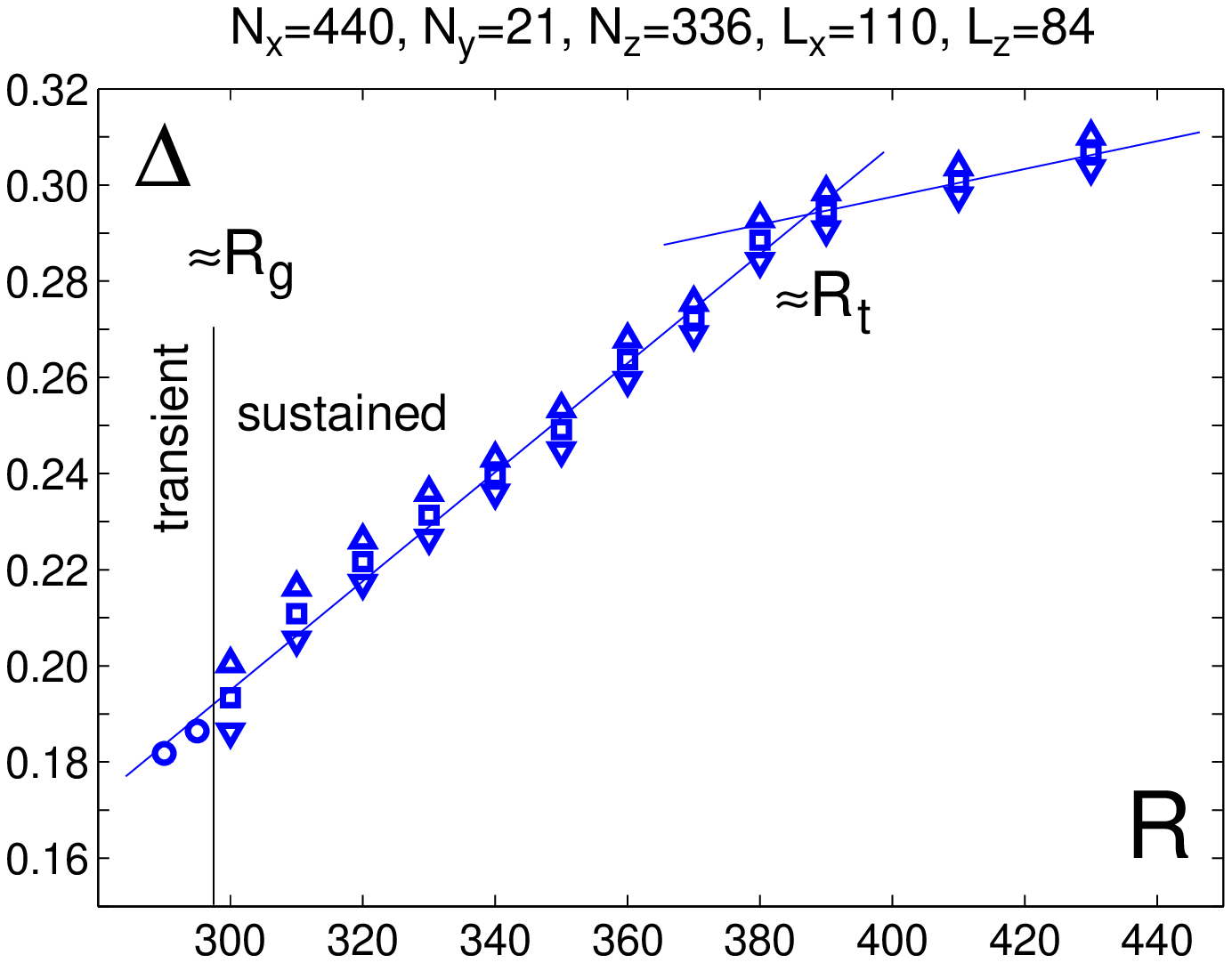}
\hspace{0.05\textwidth}
\includegraphics[width=0.45\textwidth,height=0.34\textwidth,clip]%
{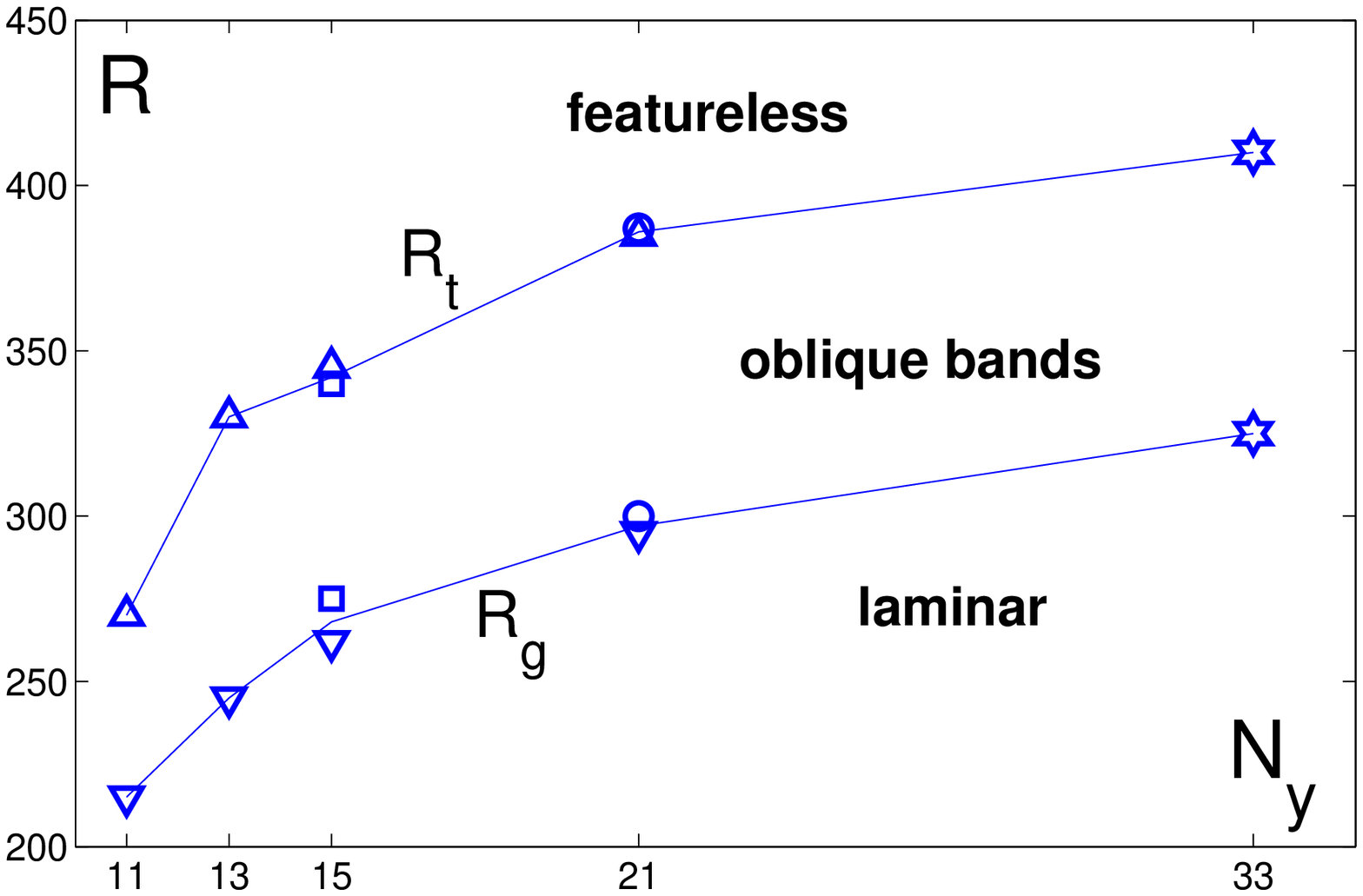}
\caption{Left: Time-averaged quantity $\Delta$ as a function of $R$
for $N_y=21$ with $(L_x,L_z)=(110,84)$ and  $(N_x,N_z)=(440,336)$.
Right: Thresholds $R_{\rm g}$ and $R_{\rm t}$ as functions of the number
$N_y$ of Chebyshev polynomials for different in-plane resolutions and domain sizes, see text.
\label{f6}}
\end{figure}

The result of the experiment for $N_y=21$, $(L_x,L_z)=(110,84)$ and
$(N_x,N_z)=(440,336)$, shown in Figure~\ref{f6} (left), is typical.
This figure displays, as  functions of $R$, the time averages of $\Delta$
as squares and the standard deviations of fluctuations as up/down
triangles. (Estimates for transient states observed below $R_{\rm g}$
are shown as open circles, see below.)

A break in $\Delta$ as a function of $R$ is
identified as the upper threshold $R_{\rm t}$. Visually, this break
corresponds to the appearance of regions where the turbulence intensity
is depleted. A continuous diagonal band forms as $R$ is decreased below
$R_{\rm t}$, which, once periodically continued in the spanwise and
streamwise directions features the expected oblique pattern.
The decrease of $\Delta$ which measures the average distance to the
laminar profile is easily understood as the result of a decrease of the
turbulent fraction, the ratio of the surface of the region still turbulent
to the total surface of the system. This fraction is determined by
thresholding the local mean perturbation kinetic energy
$\bar E_{\rm t}(x,z;t)$. In
their simulations, Barkley \& Tuckerman \cite{BT05-07} observe that the
turbulent intensity is larger inside turbulent bands than in the
featureless regime. Quantity $\Delta$ being an average over the whole
domain, by compensation this could lead to an underestimation of
$R_{\rm t}$. Rather than trying to correct for this intensification
effect%
\footnote{Incidentally, turbulence intensification warrants study in
domains of arbitrary shapes compatible with the band pattern. This
phenomenon might indeed be specific to the narrow oblique domains used
in~\cite{BT05-07}, since the periodic boundary forcing at short distance
interferes with the large-scale streamwise coherence of the streaks.}
we choose to {\it define\/} $R_{\rm t}$ as given by the position of the break further confirmed by visual inspection of the flow pattern.

The second limit $R_{\rm g}$ is determined as the value of $R$ below
which turbulence decays in less than 10000 time units. This value is
obtained from a single experiment, but a rapid change within a small
interval in $R$ is observed between sustained banded
turbulence with limited fluctuations of $\Delta$ and a turbulent regime
bound to decay which displays large excursions toward values of $\Delta$
well below the average. Below $R_{\rm g}$, taking the mean of $\Delta$
during the plateau that precedes the final decay produces the measurements
displayed as open circles which appear to be well aligned with the
other points corresponding to sustained turbulence.

At any rate our aim here is not to perform a detailed statistical study of
transient turbulence as in previous studies \cite{Eetal08} but to locate
$R_{\rm t}$ and $R_{\rm g}$ approximately as a function of the resolution.
A very conservative estimate of the precision with which
$R_{\rm t}$ and $R_{\rm g}$ are determined is $\pm5$, which meets our
purpose.

The results of the most systematic study with $(L_x,L_z)=(128,64)$,
$(N_x,N_z)=(512,256)$
are displayed in Figure~\ref{f6} (right) as up-triangles for $R_{\rm t}$,
down-triangles for $R_{\rm g}$. Values at $N_y=33$, marked with hexagrams,
are taken from the literature~\cite{Detal10}, in close agreement with
experimental observations~\cite{Petal03}. The bifurcation diagram shows
a regular shift of the interval in $R$ where bands are observed. 
These results do not depend on the precise size of the domain provided
that it is sufficiently large to accommodate a band,
as seen from a comparison of the results for $L_x=128$ and $L_z=64$
with those from the experiment with $L_x=110$ and $L_z=84$ described above
and reported as open circles at $N_y=21$. They are
also not so sensitive to the in-plane resolution, as seen
from the results at $N_y=15$ of the `high'-resolution experiment with
$(N_x,N_z)=4\times(L_x,L_z)$ compared to those of a `low'-resolution
control experiment with $N_x=L_x$ and $N_z=3L_z$ displayed as open
squares. A similar shift
was observed by Willis and Kerswell in the pipe flow case though in the
opposite direction of increased thresholds \cite{WK09}. However, in their
case, the resolution decrease was in the azimuthal  direction, which
corresponds to our spanwise direction. This difference is likely
to play a role on the turbulence sustenance mechanisms, so that no
definite inference can be made from this observation.  

Corroborating the observation of an anomalous behaviour of the fully turbulent state for $N_y=7$ and $9$ (Fig.~\ref{f1}, left;
Fig.~\ref{f2}, top-left), banded patterns
are not obtained in these cases. 
On the contrary, a continuous decrease of $\Delta$ is observed as $R$ is
decreased. For $N_y=9$, a featureless turbulent state similar to
that at $R=450$ (Figure~\ref{f2}, top-left) can indeed be maintained
as low as $R=185$ while for smaller $R$, a small-scale streamwise structure happens to grow on top of a large scale spanwise modulation of the turbulence intensity. This manifestly non-physical, slowly time-dependent,
numerical solution finally decays abruptly as the Reynolds number is
further decreased but a detailed study of this transition is
pointless. Similarly, an even more exotic but equally uninteresting
chaotic state is obtained for $N_y=7$, and maintained at even lower $R$.

\begin{figure*}
\begin{minipage}{0.49\textwidth}
\begin{center}
local mean perturbation energy\\
\includegraphics[width=0.9\textwidth,clip]{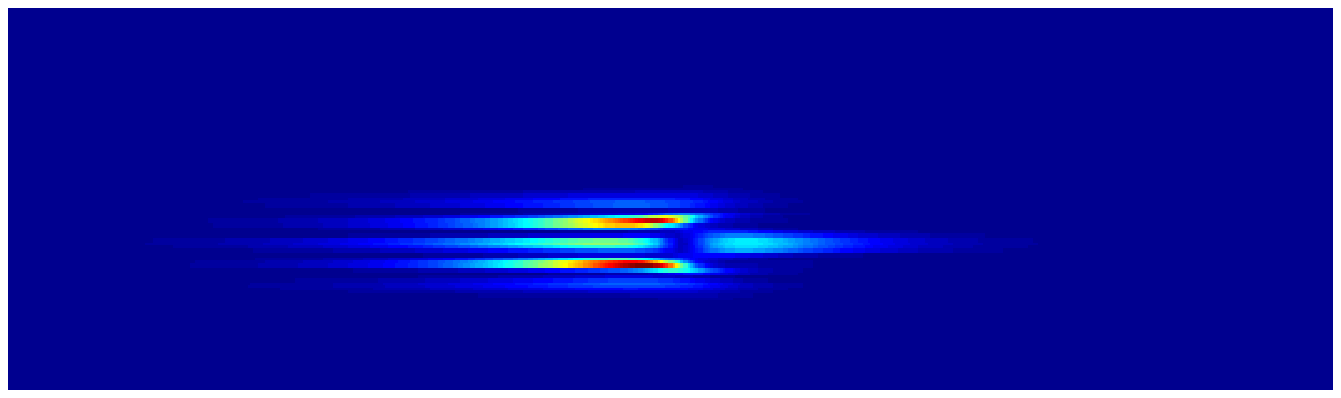}\\[2ex]
streamwise perturbation velocity component $u$\\[1ex]
\includegraphics[width=0.9\textwidth,clip]{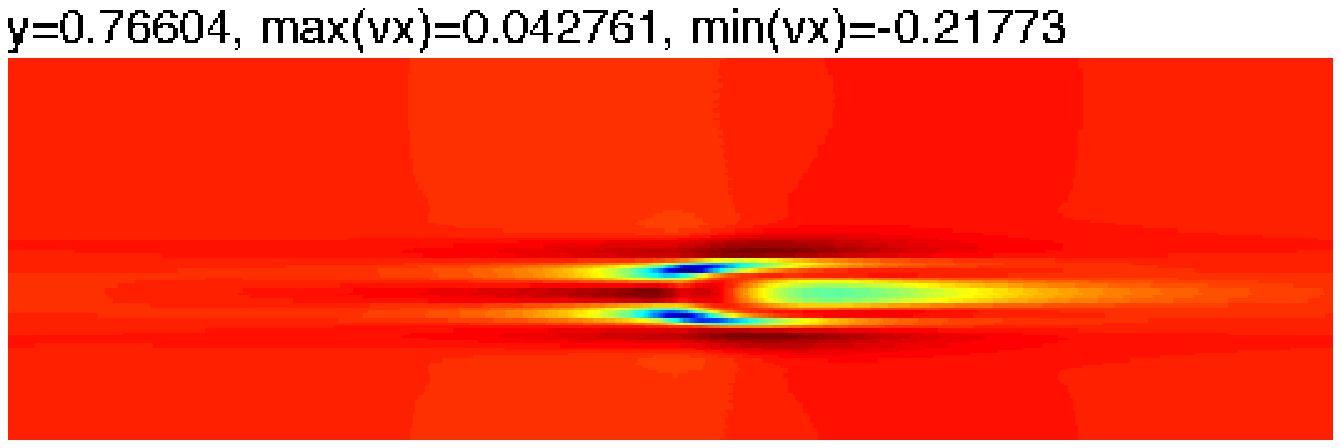}\\
\includegraphics[width=0.9\textwidth,clip]{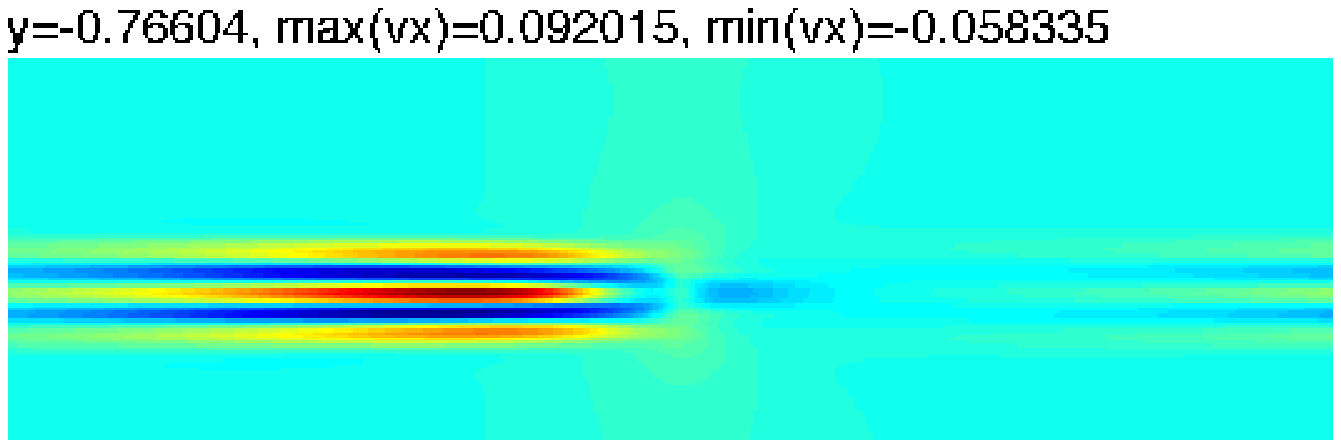}
\end{center}
\end{minipage}\hfill
\begin{minipage}{0.45\textwidth}
\includegraphics[width=\textwidth,clip]{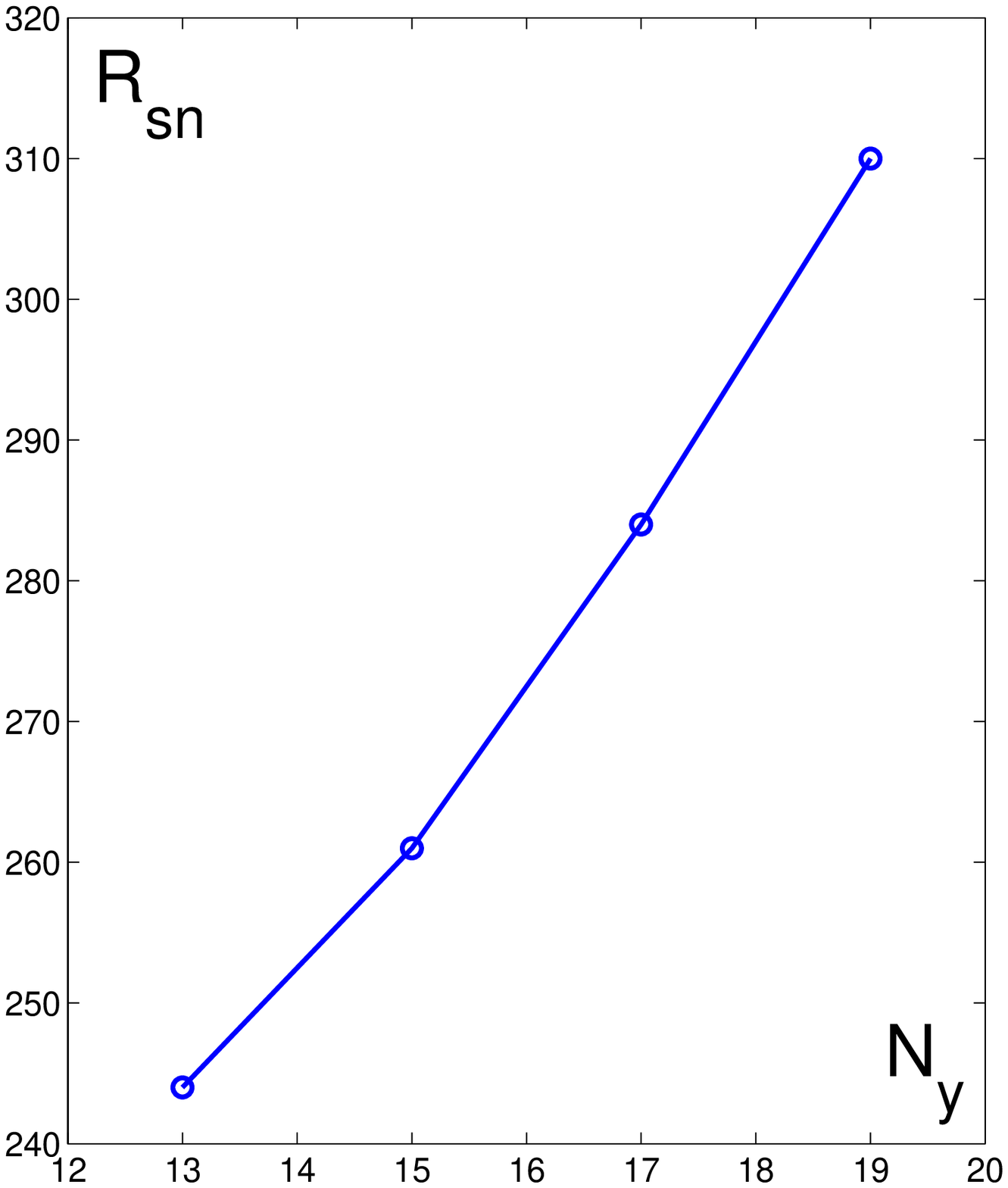}
\end{minipage}
\caption{Left: Spurious `edge state'-like numerical solution for $N_y=19$
close to the saddle-node bifurcation point $R\approx309.5$ ($L_x=110$,
$L_z=32$, $N_x=440$, $N_z=128$); this solution is mostly concentrated
in the upper half of the channel ($y>0$) as understood from the comparison
of the variation ranges of $vx\equiv u(x,y,z,t)$ for $y=+0.766$ and $y=-0.766$ at steady state ($t\to\infty$); Right: Variation of the
saddle-node threshold $R_{\rm sn}$ as
a function of $N_y$ for the family of spurious numerical solutions.
\label{spurious}}
\end{figure*}

A caveat is however required: at moderate (i.e. not the lowest) resolution,
spurious stable nontrivial nearly-steady states can be
found at values of $R$ in the lower part of the range where the bands
exist and below, in the form of localised states resembling edge states,
e.g. \cite{Detal09}.
An example is shown in Figure~\ref{spurious} (left).
As understood from the display of the local mean perturbation energy
and the streamwise component of the perturbation velocity, this solution
has broken the symmetries of the plane Couette flow since it is
predominantly localised in $y\in[0,1]$ and not symmetrically
distributed over $[-1,1]$.
As a result of the symmetry breaking, this localised state moves slowly
in the streamwise direction. The resolution being given, such a state can
be viewed as a solution to some dynamical system and, so, belonging to
some solution branch along which it can bifurcate as $R$ is varied.
It turns out that this state exists in a limited
range between a lower saddle-node threshold where it disappears and
an upper threshold where it experiences a Hopf bifurcation, and next
nucleates a turbulent spot, which is expected and thus not studied in
detail. In fact, this solution belongs to a family that exists for
a range of cross-stream resolutions $N_y$. It was obtained from another
one obtained during the decay of a turbulent state for $N_y=15$ and
$R=270$, then carried to $N_y=13$ on the one hand, and to $N_y=17$ and
next $N_y=19$  on the other hand, but could be stabilised neither for
$N_y=11$ nor for $N_y=21$. Figure~\ref{spurious} (right) displays the
threshold of the saddle-node bifurcation through which this solution
family disappears.
This threshold is seen to increase rapidly and to lie below $R_{\rm g}$
for $N_y=13$ or 15, and above for $N_y=19$, which is probably the
reason why we cannot find it for $N_y=21$ without the help of a sophisticated continuation procedure. This family of spurious solutions
is most probably not unique. Unlike genuine edge states that
are unstable by construction and proposed to represent gates on the
boundary of attraction basin of the base flow, they are stable at
least within some range of Reynolds numbers; the mechanisms by
which they are maintained should however be essentially identical.%
\footnote{Though we were unable to find any report on the existence of
similar spurious numerical solutions---fragile with respect to a
resolution change while fitting the framework of conventional bifurcation
theory---in the open literature, we heard from Y.~Duguet that
the problem also arose in the search for exact solutions in MFU-sized systems by Schmiegel (1999) and Gibson {\it et al.} more recently.}
We stress that this unwanted side effect of under-resolution here
acts on nontrivial but non-chaotic states and is not expected to spoil
our study of turbulent regimes which are sensitive to noise inherent in chaos but statistically robust enough to withstand perturbations due to
truncation errors.

To conclude this section, in spite of the caveat above, evidence has been
given that as long as unsteady (turbulent) states are considered,
reducing the wall-normal resolution is a viable modelling strategy.
Of course the best possible resolution is desirable but, in view of a
quantitative reproduction of the transitional range of plane Couette
flow, we can recommend that the number $N_y$ of Chebyshev polynomials be
kept larger than or equal to 11, whereas a tolerable shift of the upper
and lower bounds of that range is obtained provided that $N_y\ge15$.
In-plane resolutions with $N_x\ge L_x$ and $N_z\ge 3L_z$ also appear
satisfactory, with corresponding numbers of Fourier modes
$\frac23 N_{x,z}$.

\section{Discussion}

We begin by presenting preliminary results obtained in parallel
with the study presented above to illustrate how it can be used, before
making more general comments on the numerical approach to transitional
wall-bounded flows.

\begin{figure}
\begin{center}
\includegraphics[width=0.55\textwidth,clip]{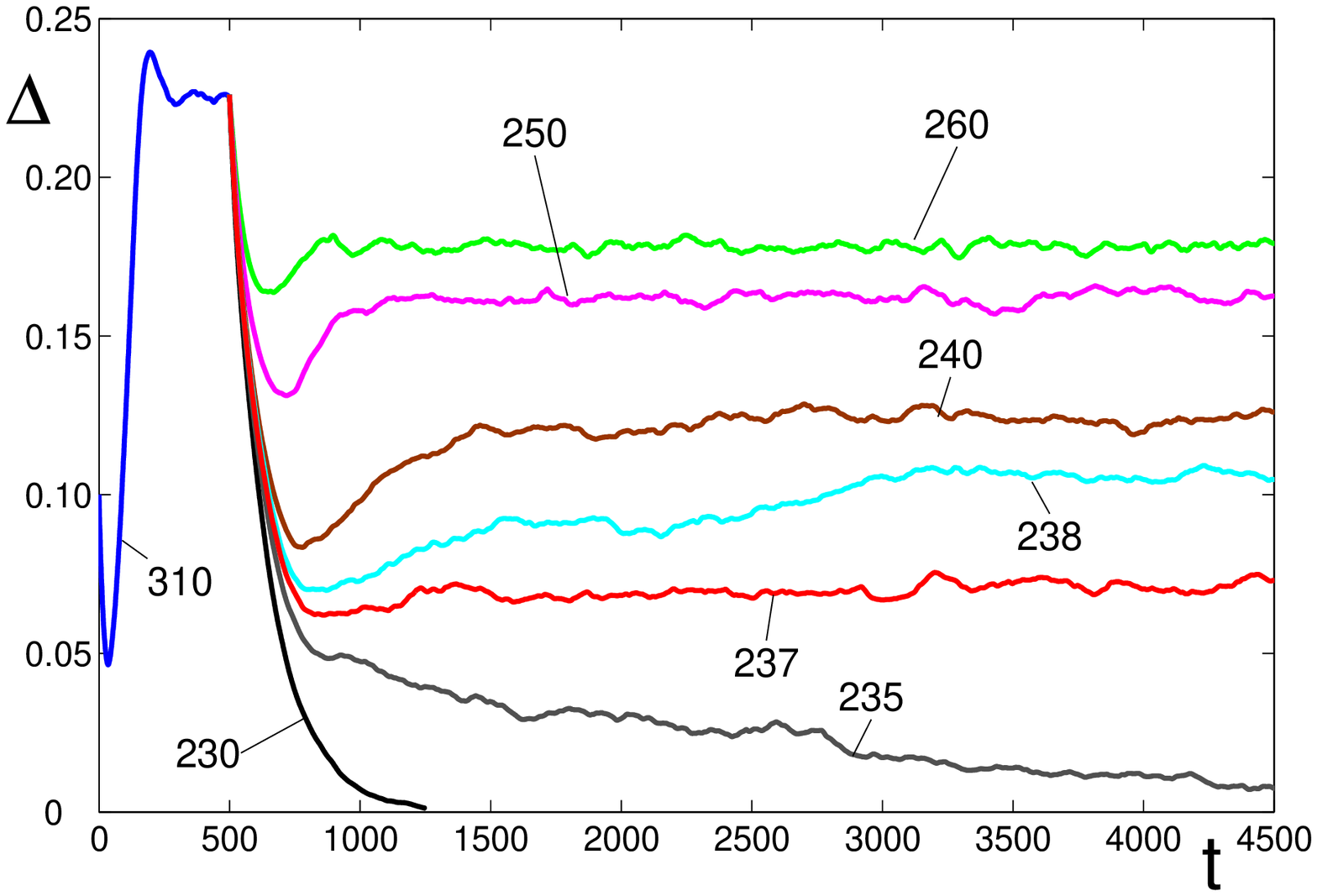}\\[2ex]
\includegraphics[width=0.45\textwidth,clip]{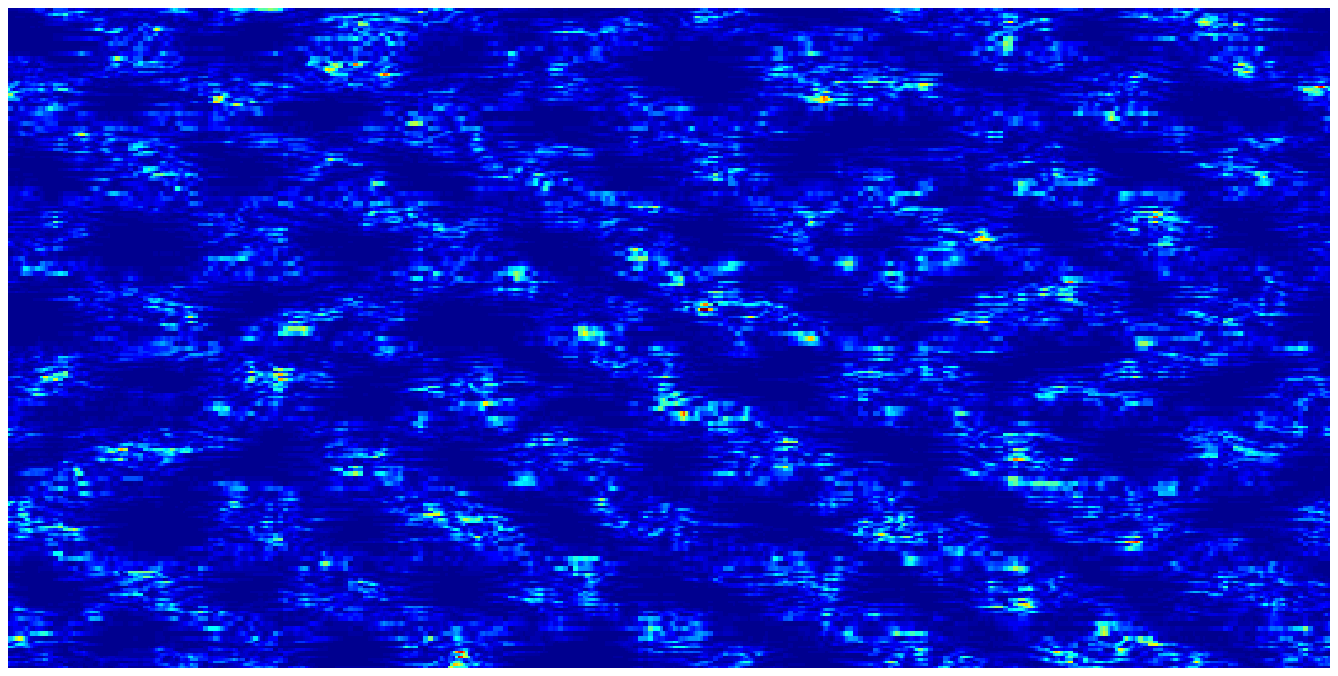}
\hspace{0.05\textwidth}
\includegraphics[width=0.45\textwidth,clip]{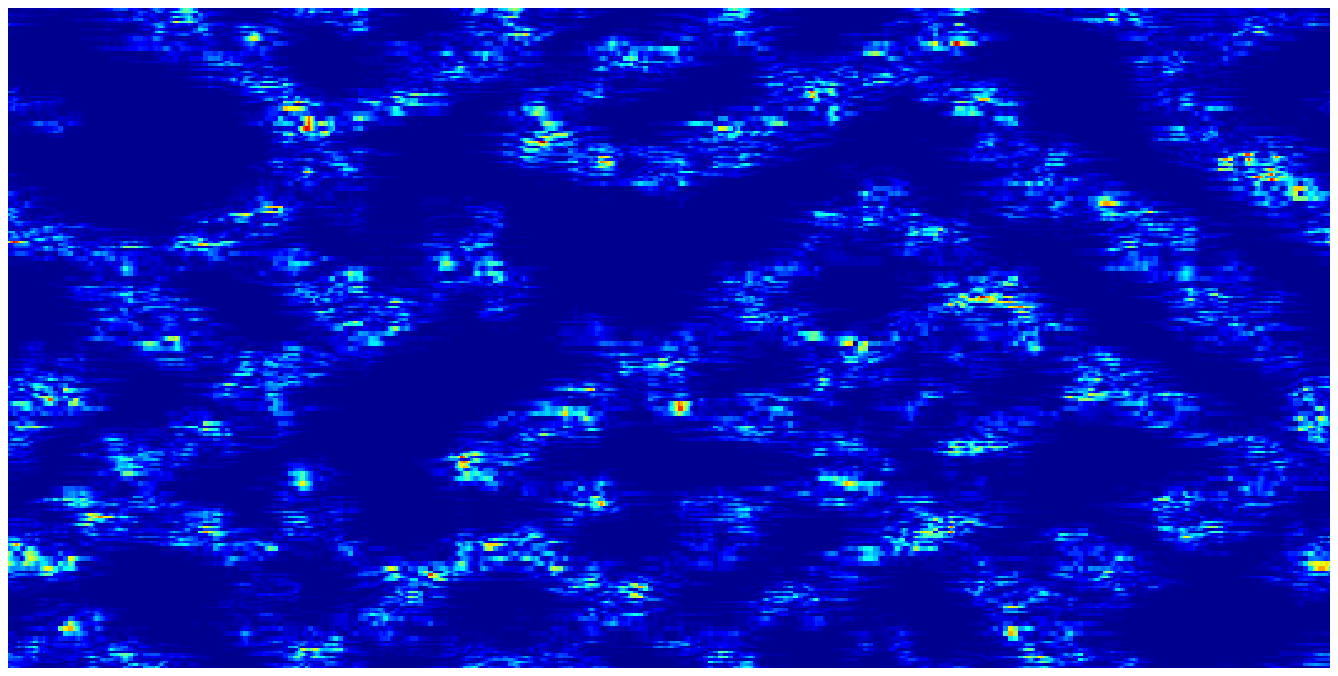}\\[2ex]
\includegraphics[width=0.45\textwidth,clip]{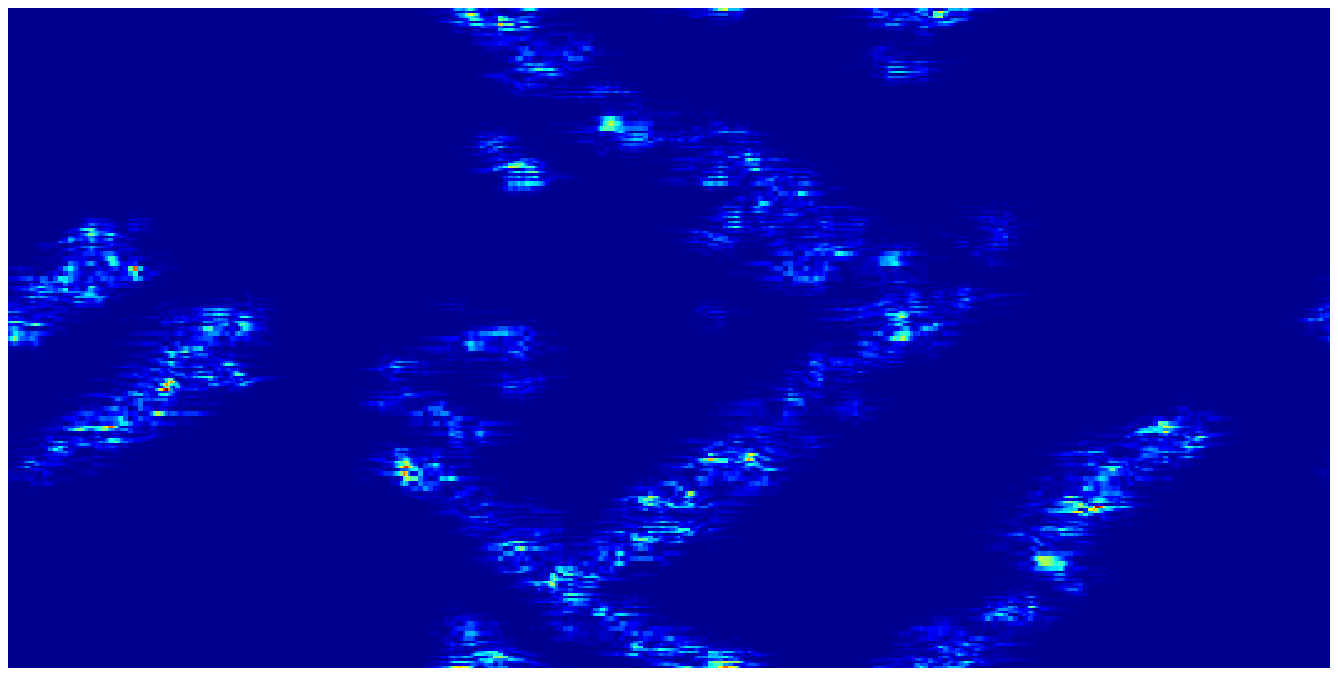}
\hspace{0.05\textwidth}
\includegraphics[width=0.45\textwidth,clip]{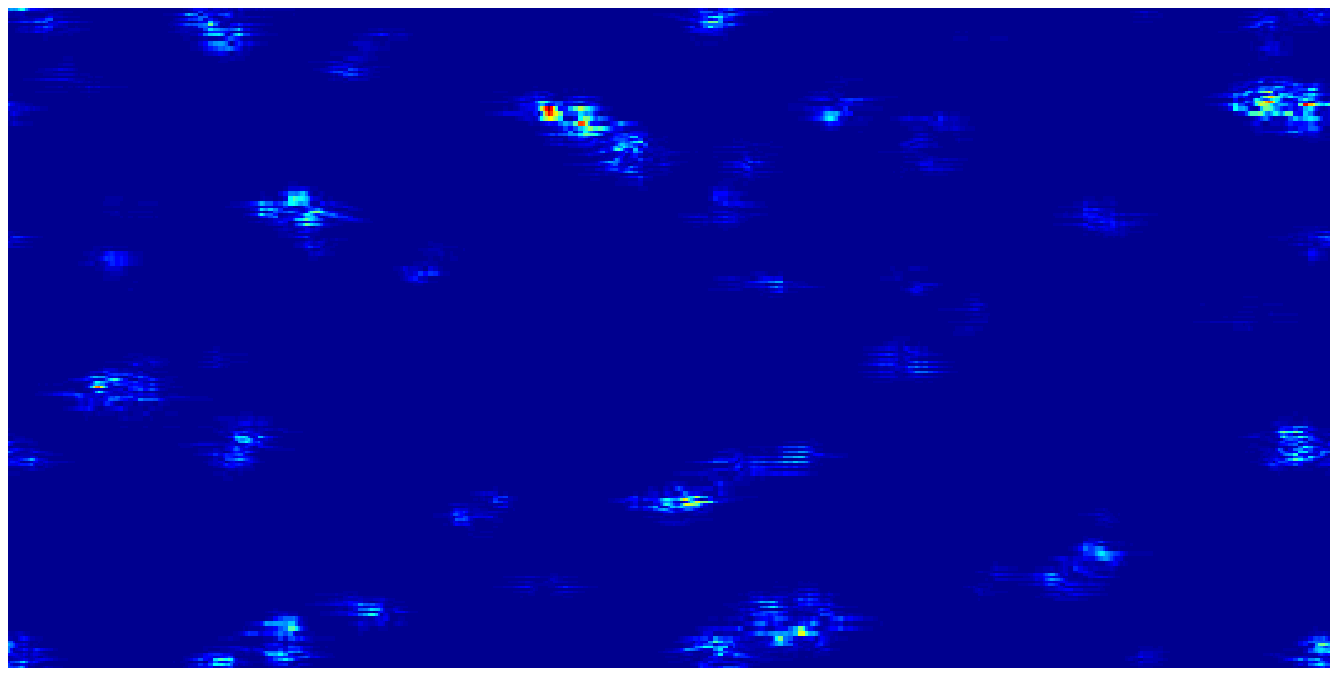}
\end{center}
\caption{Quench experiment for $N_y=11$,
 $N_x=L_x=682$, $N_z=L_z=341$.
Top: Time series of quantity $\Delta$ for the values of $R$ indicated.
Bottom, from left to right and top to bottom: Snapshots of pattern
obtained in the different asymptotic regimes obtained at $t=4000$ after
the quench for $R=250$ (continuous bands),
$R=240$ (interrupted bands), $R=237$ (border state),
and at $t=500$ for $R=235$ (decaying spots).\label{fex_st} }
\end{figure}

\paragraph{$\bullet$ Extreme case.} The emergence of turbulent bands
first appears to be an extremely robust phenomenon. Here we show 
simulation results in the most extreme conditions that we have considered,
namely $N_y=11$ and $N_x=L_x$,
$N_z=L_z$. The lateral size of the domain is $L_x=682$ and
$L_z=381$, which is already larger than in the early Saclay
experiments~\cite{BC98} and of the same order of magnitude as
in the latest experiments~\cite{Petal03} or the simulations reported
in~\cite{Detal10} but with the computation power of a desk-top computer.%
\footnote{Linux operated Dell computer with Intel Core2 DUO CPU E8400
3.00~GHz, 4 Go DDR2 memory.}
From results in Fig.~\ref{f6} (right), we expect $R_{\rm t}\simeq270$ and
$R_{\rm g}\simeq215$. Quench experiments similar to those of
Bottin \cite{BC98} are performed. Results are displayed in 
Figure~\ref{fex_st}.
A turbulent state is prepared
at $R=310\gg R_{\rm t}$. At time $t=500$, $R$ is suddenly reduced to some
final value $R_{\rm f}$. Time series of the quantity $\Delta$ are shown
in the top panel. Snapshots of the solution are taken at regular times.
The experiment is stopped at $t=4000$ after the quench or when the
laminar state is recovered.

After the quench, a fast drop of $\Delta$ is observed with a pronounced
undershoot for $R=260$, $250$, and $240$. During the drop, the evolution
amounts to a general, more or less uniform, breakdown of the turbulent
state that, when the minimum is reached, leaves a few turbulent patches
of limited extent. The turbulent fraction then re-increases and bands
form. For $R=260$ and $250$ continuous bands are obtained but, in
contrast with experiments where a single wave vector
$\mathbf{k}^{\rm patt}$ was selected, here two wavevectors symmetrical
with respect to the streamwise direction ($\pm k_z$) dominate the
pattern as displayed in the top-left image. The 
quench at $R_{\rm f}=240$ ends with a regime where bands are fragmented
in the top-right image. For smaller $R_{\rm f}$, the undershoot is less
pronounced and the recovery stage slower. The asymptotic regime reached at
$R=237$ will be called a {\it border state\/} since---within
the quench protocol---it sits on the frontier separating decaying
from sustained regimes. As such, arguably, it could have been called an
`edge state' but this term has been already extensively
used in the study of the transition in terms of dynamical systems
\cite{Eetal08} which focuses of the temporal dynamics of localised
structures that are {\it exceptional\/} limit sets. In contrast, the
state shown in the bottom-left image appears {\it typical\/} of the
spatiotemporal aspects of the transition.
For $R_{\rm f}=230$, the system does not recover
and the turbulent patches decay immediately. Apart from an important
downward shift of $R_{\rm t}$ and $R_{\rm g}$, all of the observed scenario
is identical to results described by Duguet {\it et al.} \cite{Detal10}
obtained in a much better resolved context, or to Prigent's experiments
\cite{Petal03}.

From the above one could conclude that $R_{\rm g}\approx237$, which
is somewhat larger than the value obtained using the adiabatic protocol
used to obtain Figure~\ref{f6} (right) in which one gets $R_{\rm g}\approx215$. This discrepancy might be due to the difference in the
in-plane resolution: $N_{x,z}=4L_{x,z}$ for results in that figure
vs. $N_{x,z}=L_{x,z}$ here. A similar effect was already apparent for
$N_y=15$ and the two resolutions considered: $R_{\rm g}\simeq262$ was
obtained with $N_{x,z}=4L_{x,z}$ (down triangle) and
$R_{\rm g}\simeq275$ with $N_x=L_x$, $N_z=3L_z$ (square),
which is slightly better than now ($N_z=L_z$). Another, more likely,
explanation can also be found in the fact that the threshold obtained
in the quench experiment is only an upper bound to $R_{\rm g}$:
The quench protocol always involves a recovery stage starting at
the end of the undershoot. The corresponding state is an assembly of
turbulent spots similar to what is pictured in figure~\ref{fex_st}
(bottom, right) during the early decay at $R=235$. In this respect,
the quench experiment is an initial value problem similar to
that of Duguet {\it et al.\/} \cite{Detal10}, or to any experiment
in which spots are triggered.
In contrast, the adiabatic decrease of $R$ is the sole protocol supposed
to yield, by continuation,
the global stability threshold defined as the lowest value
of $R$ above which sustained turbulence has a non-empty attraction
basin. Except in a well-conducted adiabatic experiment, this
attraction basin may indeed be hard to find until one reaches
values of $R$ where it has a sizeable breadth, which is
somewhat beyond $R_{\rm g}$.

The superposition of two sets of bands illustrated above instead of
one, as expected from the experiments, is also the result of a resolution
which, though qualitatively satisfactory, is nevertheless too poor since
ongoing simulations with $N_y=15$ show a single dominant wavevector in
the range of Reynolds numbers corresponding to bands. Interpreting
the result in terms of pattern formation and Ginzburg--Landau envelope equations \cite{Petal03}, when the
resolution is low ($N_y=11$) the coefficient of the
cubic term $|A_{\mp}|^2 A_{\pm}$ accounting for the interaction
between wavevectors $+k_x$ and $-k_x$ (explicitly defined in the next
paragraph) is under-estimated compared to the coefficient of the term
$|A_{\pm}|^2 A_{\pm}$ accounting for self-interaction, so that
each orientation can develop to form a rhombic pattern;
a ratio compatible with experimental findings is restored
by increasing the resolution, ending in a stripe pattern ($N_y=15$). 
On the other hand, it should be stressed that, despite the
constraints brought by the periodic boundary conditions, the
pattern's wavelengths and their variation are correctly reproduced
since we can measure
$\lambda_x\approx114=L_x/6$ (and not $L_x/5=136$ nor $L_x/7=97$)
for both $R=250$ and $240$ and $\lambda_z=68\>(=L_z/5)$ and
$85\>(=L_z/4)$ for $R=250$ and $240$, respectively. 

\begin{figure}
\begin{center}
\includegraphics[angle=90,width=0.24\textwidth,clip]{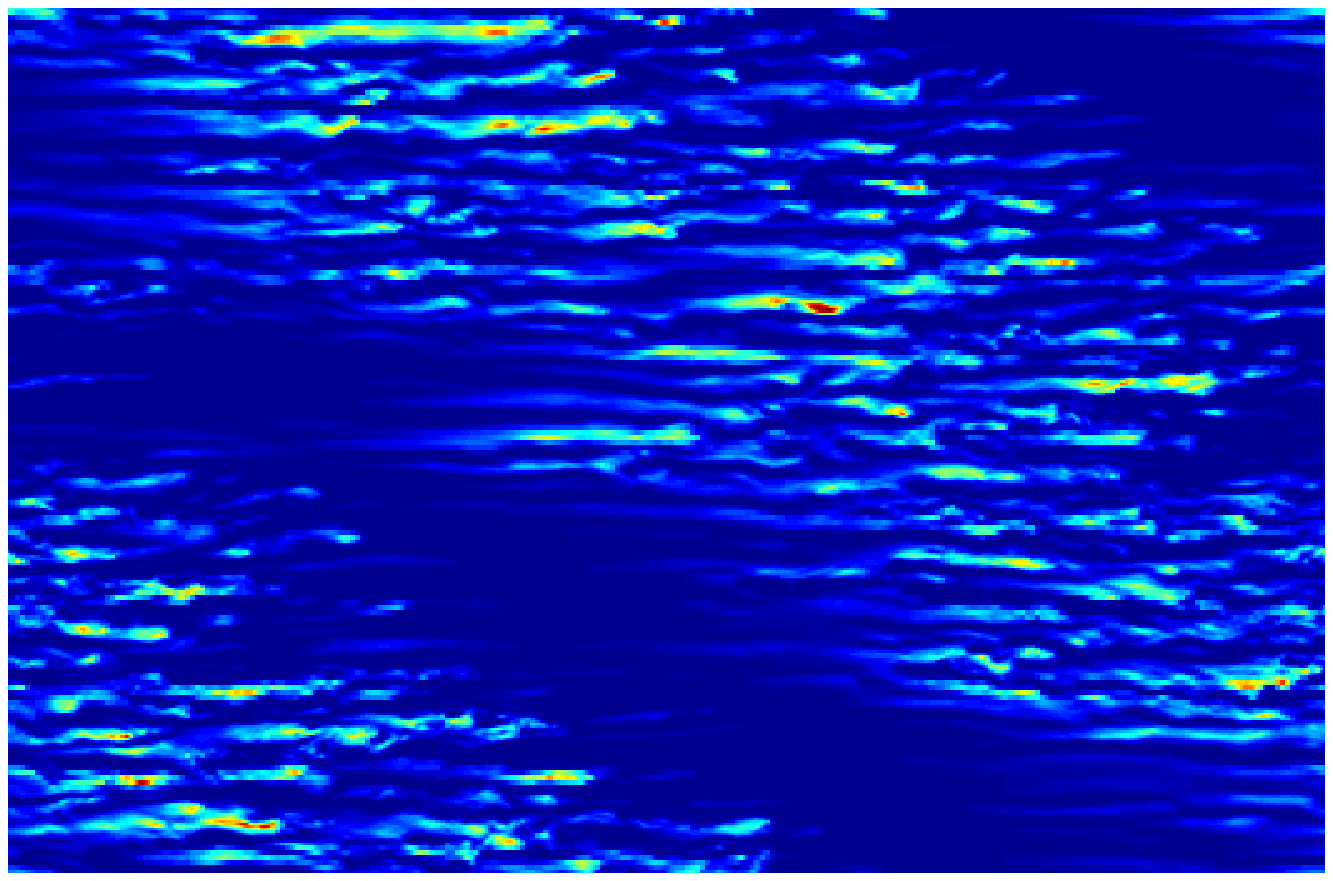}\hfill
\includegraphics[angle=90,width=0.24\textwidth,clip]{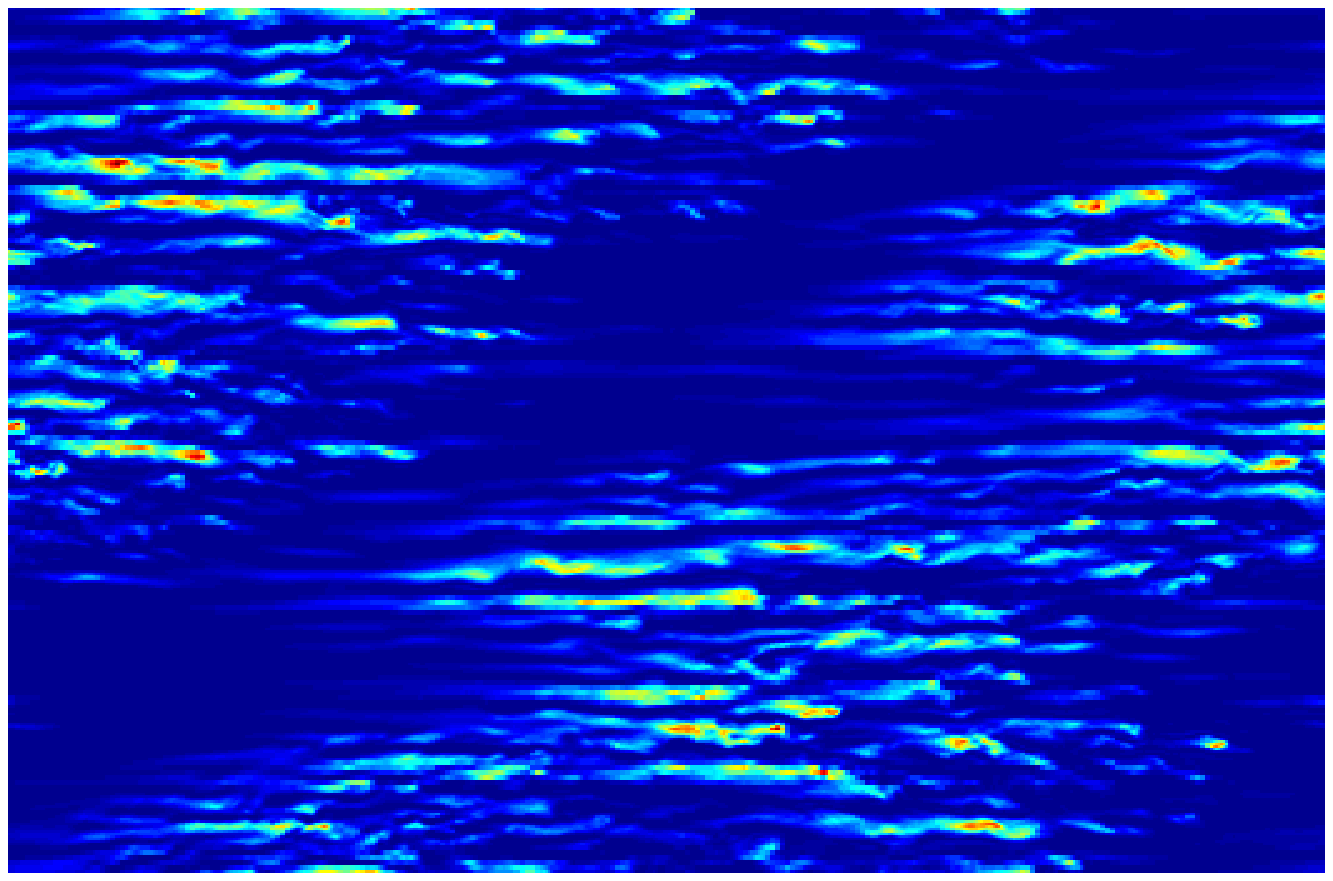}\hfill
\includegraphics[angle=90,width=0.24\textwidth,clip]{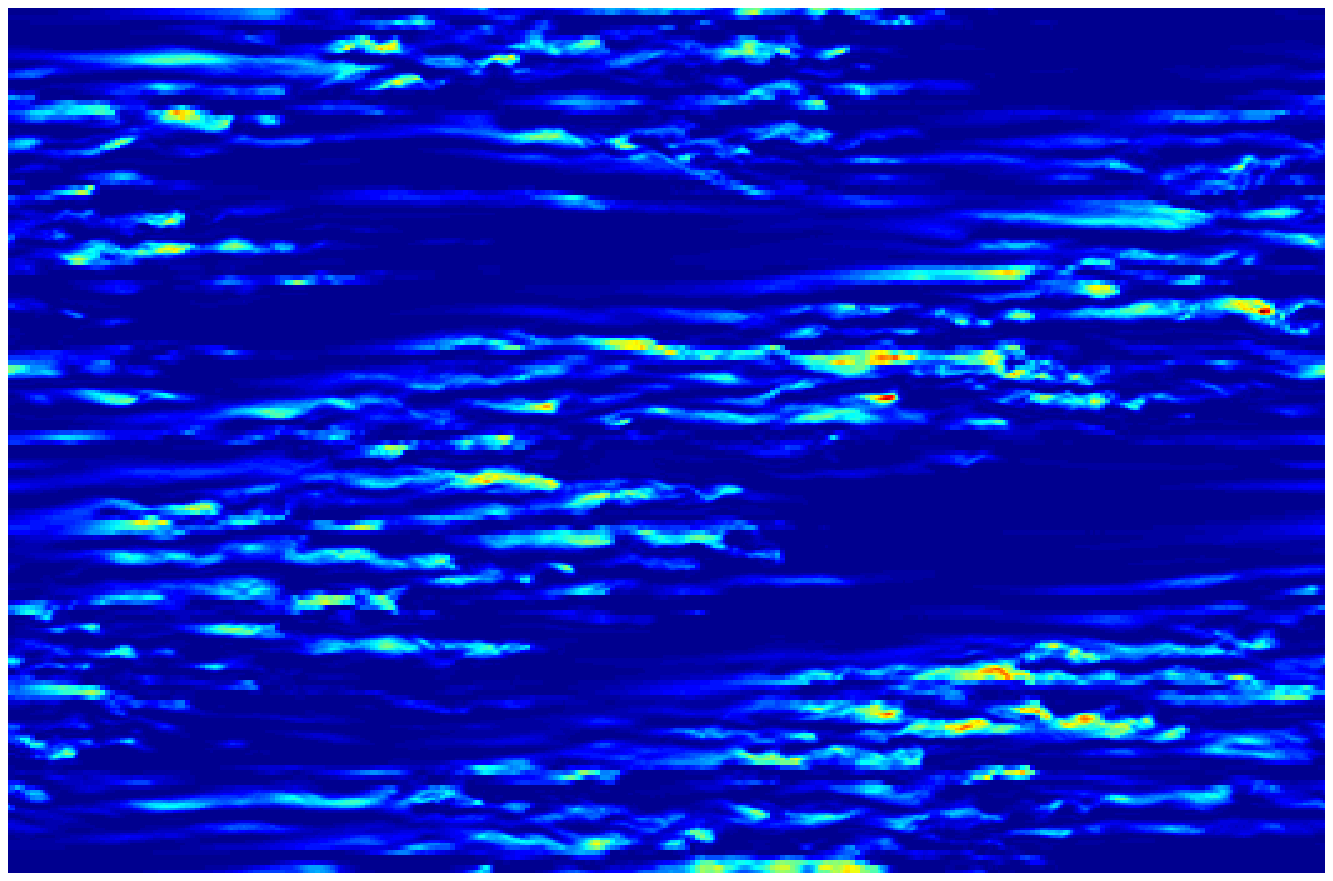}\hfill
\includegraphics[angle=90,width=0.24\textwidth,clip]{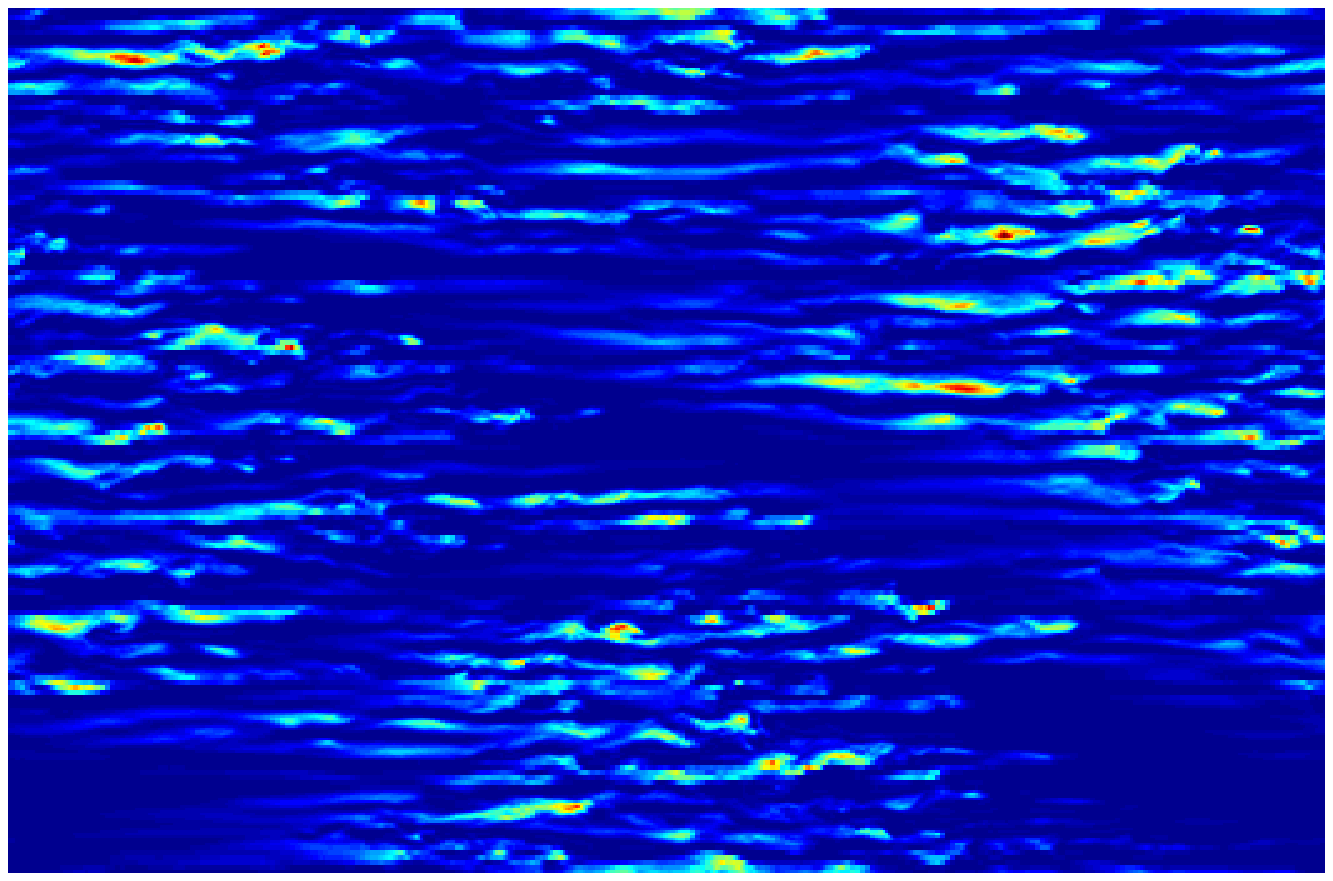}\\[4ex]
\includegraphics[width=0.8\textwidth,clip]{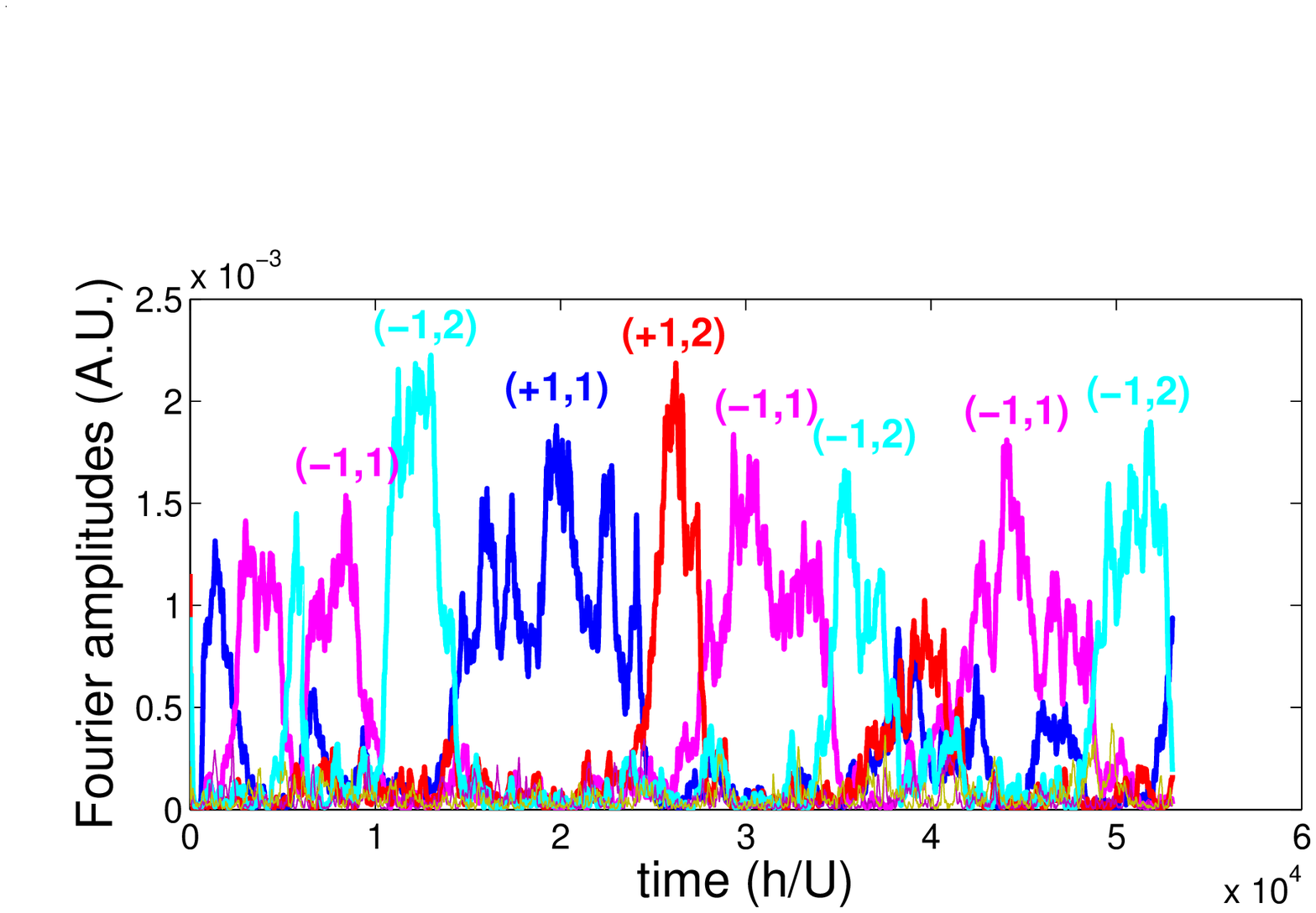}
\end{center}
\caption{\label{fop} Four snapshots of the solution for $R=315$
($N_y=15$, $L_x=110$, $N_x=512$, $L_z=84$, $N_z=256$, $x$ axis is
vertical). Bottom: Time series of the amplitude of the different
dominant Fourier modes.}
\end{figure}

\paragraph{$\bullet$ A better resolved experiment.} The case considered above stays too close to the boundary of the basin where DNS faithfully accounts for transitional plane Couette flow. The real motivation of our
work is to consider less extreme simulation conditions in view of a quantitatively better rendering of the different regimes observed.
With our present computational capabilities, this can only be
done by increasing $N_y$ while simultaneously keeping a sufficiently
high in-plane resolution at the expense of decreasing the size of the domain. Accordingly we have considered a system of size sufficient to
afford at least a full elementary cell of the experimentally observed pattern, i.e.  $L_x\sim \lambda_x$ and $L_z\sim\lambda_z$. The example 
shown here is with $L_x=110$ and $L_z=84$ and a resolution $N_y=15$, $N_x=512$, $N_z=256$. It aims at the identification of an appropriate
order parameter for the pattern \cite{RM09}. The top row of
Figure~\ref{fop}
displays snapshots of the local mean energy at different
times during a long simulation at $R=315$. Band patterns
with a single band leaning to the right or to the left, with two bands leaning to the left, or a messy situation, can be observed.
Fourier analysis then yields either one dominant mode (well formed
patterns, Fig.~\ref{fop}, left and centre) or several
interacting modes (defective pattern, Fig.~\ref{fop}, right).
In the present context, Fourier amplitudes are good candidates to play
the role of order parameters. They were indeed considered by Barkley
{\it et al.} in \cite{BT05-07} and \cite{TBD09}.
The time series of the Fourier amplitudes shown as functions of time
in Figure~\ref{fop} (bottom) indicate an apparently random alternation
of states that can be characterised by a dominant mode ($\pm$ indicate
left or right, and 1 or 2 the number of bands, i.e. $k_x=\pm 2\pi n_x/L_x$
and $k_z=2\pi n_z/L_z$, with $n_x=1$ and $n_z=1,2$). Preliminary
analysis suggests that the sojourn times in one or another state are
exponentially distributed, which suggests an approach in terms of a
stochastic multi-well process. This interpretation follows from the
probable existence of a Landau potential with minima appropriate
to describe the several possible pattern configurations, and the presence
of a permanent excitation due to the noise generated by
turbulence~\cite{Petal03}. But, to be validated, such an interpretation
requires large amounts of data still being gathered in different
conditions in order to get the variation of the order parameter, its mean
value and standard deviation, as functions of $R$ for different
aspect-ratios $L_{x,z}$.

\paragraph{$\bullet$ Final remarks.}
When combined with the reliability assessment provided in Sections~\ref{s2}
and \ref{s3}, these preliminary results bring an interesting contribution
to the phenomenology of plane Couette flow at minimal numerical cost.
The pattern's main characteristics are recovered. Turbulent band
formation appears to be a robust feature in the transitional range
$R\>\in\>[R_{\rm g},R_{\rm t}]$. The order of magnitude of the
pattern's wavelengths and their variations with $R$ are also well
reproduced in the simulations, even at the lowest possible resolution.
The price to be paid for the resolution decrease just seems to be a
regular shift of the transitional range toward lower Reynolds numbers.

These results also illustrate the specifically {\it spatiotemporal\/}
features of the transitional range. In particular, turbulence decay may
not well be rendered  by the {\it temporal\/} approach implemented
in low-dimensional dynamical systems theory (chaotic
transients~\cite{Eetal08}). The latter approach remains adapted
to chaotic but spatially coherent dynamics in domains a few MFUs wide.
It can be of use to understand the nucleation of laminar patches
of limited extent within featureless turbulence. It is however
unable to account for the regular regression of turbulent domains
coexisting with laminar flow which marks the decay stage in large aspect ratio systems, either in the laboratory or in the computer. This is
attested by the variation of $\Delta$ shown here for $R=235$ and $N_y=11$
in Fig.~\ref{fex_st}, but typical of better resolved cases.

Interesting results have already be obtained in an elongated
inclined domain~\cite{BT05-07,TBD09} but confinement by periodic
boundary conditions in the direction parallel to the short side 
perturbs the long-range streamwise coherence of the streaks
(Fig.~\ref{f2}). This warrants further
study and leads us to suggest that one should consider domains that
are at least as large as one elementary cell $(\lambda_x,\lambda_z)$
of the band pattern.%
\footnote{This option was taken by Barkley in his latest work on band
formation presented at the conference {\it New Trends On Growth And
Form\/}, Agay (France) June 20-25, 2010.}

As a modelling strategy within the extended systems perspective, results
presented here drastically improves over the model previously elaborated
in \cite{LM07} and used in \cite{Ma09}. Though that model is amenable to
analytic treatment in view of the elucidation of the mechanisms producing
the experimentally observed turbulence modulation, the present
numerical findings point out the role of its effective cross-stream
resolution which is much too low. On the other hand, the fact that
limited resolution reproduces the main features of the dynamics of the
flow in the transitional range suggests that the observed pattern
formation does not involve processes taking place in a thin boundary
layer close to the plates where high-order cross-stream modes are of importance, but larger scale interactions in the bulk of the shear
where structures controlled by moderate-order modes operate
(see \cite{SK09} for the companion problem of puff sustainment in
pipe flow). This observation could motivate the search for a model
of intermediate complexity along the lines traced in~\cite{LM07}
by pushing the Galerkin expansion a little further in view of
an analytical approach.

To conclude, plane Couette flow presents itself as an academic prototype
of wall-bounded flows, with the extreme condition that it is linearly
stable for all $R$, forcing its subcritical character. 
Low resolution simulations allows numerical experiments at minimal cost
in circumstances where dynamics in {\it physical space\/} becomes more
relevant (laminar/turbulent coexistence) than in {\it phase space\/}
where the collection of competing exact solutions becomes
increasingly large and their properties difficult to exploit. Our
unorthodox approach of decreasing the resolution in a controlled way 
may be considered as a modelling methodology which will allow refined
statistics, in view of Pomeau's thermodynamic analogy~\cite{Po86,BPV98},
and help to pose appropriate questions about the physics behind band
formation. The approach can certainly not be extended to the fully
developed regime nor to initial spot development where high resolution
is an important issue, but moderately turbulent regimes involved
in the transitional range of less academic flows could take advantage
of it when the lateral extension is of primordial interest.

Acknowledgements:
P.M. wants to thank Y.~Duguet and G.~Kawahara for interesting
discussions related to this work, and the latter for his invitation
to present it in Osaka. Helpful comments of a referee are also deeply acknowledged.

\end{document}